\documentclass[submitting]{nst}

\usepackage{subfigure,dcolumn,lipsum}
\usepackage{epstopdf}
\usepackage{mhchem}
\usepackage{upgreek}

\usepackage{color}
\usepackage[dvipsnames]{xcolor}

\usepackage[utf8]{inputenc}
\def\be{\begin{eqnarray}}
\def\ee{\end{eqnarray}}
\newcommand{\beq}{\begin{eqnarray}}
\newcommand{\eeq}{\end{eqnarray}}

\def\t{\tilde }

\def\p{{\bf p}}

\def\e{\epsilon}
\def\rmd{\rm d}

\def\P{{\mathcal P}}

\def\F{{\mathcal F}}

\def\L{{\mathcal L}}

\def\bra{\langle}
\def\ket{\rangle}

\def\ln{\mbox{ln}}

\def\Kn{\rm Kn}

\def\Eq#1{Eq.~(\ref{#1})}
\def\Eqs#1{Eqs.~(\ref{#1})}

\def\Fig#1{Fig.~\ref{#1}}
\def\Sect#1{{Section~\ref{#1}}}

\def\ie{{\it i.e.}}

\def\eg{{\it e.g.}}

\begin{document}

\title{Recent development of hydrodynamic modeling in heavy-ion collisions}

\author{Chun Shen}
\affiliation{Department of Physics and Astronomy, Wayne State University, Detroit, MI 48201, USA}
\affiliation{RIKEN BNL Research Center, Brookhaven National Laboratory, Upton, NY 11973, USA}
\email{chunshen@wayne.edu}

\author{Li Yan}
\affiliation{Key Laboratory of Nuclear Physics and Ion-Beam Application (MOE) \& Institute of Modern Physics\\
Fudan University, 220 Handan Road, 200433, Yangpu District, Shanghai, China}
\email{cliyan@fudan.edu.cn}

\begin{abstract}

We present a concise review of the recent development of relativistic hydrodynamics and its applications to heavy-ion collisions. Theoretical progress on the extended formulation of hydrodynamics towards out-of-equilibrium systems is addressed, emphasizing the so-called attractor solution. On the other hand, recent phenomenological improvements in the hydrodynamic modeling of heavy-ion collisions with respect to the ongoing Beam Energy Scan program, the quantitative characterization of transport coefficients in the three-dimensionally expanding quark-gluon plasma, the fluid description of small colliding systems, and some other interdisciplinary connections are discussed.

\end{abstract}


\maketitle


\section{Introduction}\label{sec.I}
Smashing heavy nuclei at high energies in large particle accelerators routinely creates extreme conditions to study the properties of many-body systems whose interactions are governed by Quantum Chromodynamics (QCD). Within a few yetoseconds ($10^{-24}$s), the collision systems are squeezed to $10^{30}$ atm and reach several trillion degrees of Kelvin. A novel state of matter with deconfined quarks and gluons are formed under such extreme conditions, called Quark-Gluon Plasma (QGP). 

The QGP created in laboratories is a relativistic dynamical system, which expands and evolves like an almost perfect liquid~\cite{Shuryak:2014zxa}. Size of the liquid droplet depends on size of the colliding nucleus, which may vary from $O(10)$ fm in gold-gold collisions at the Relativistic Heavy-Ion Collider (RHIC) at Brookhaven National Laboratory, or  lead-lead collisions at the Large Hadron Collider (LHC) at CERN, to $O(1)$ fm in small colliding systems such as the proton-lead or even proton-proton collisions carried out at these facilities. Fluidity of QGP is one of the major subjects that has been explored heavy-ion collisions. From experiments, it is analyzed extensively through various types of long-range multi-particle correlations of the observed hadrons, known as the signatures of collective flow~\cite{Ollitrault:1992bk,Alver:2010gr,Voloshin:2011mx}. Theoretical model calculations using relativistic viscous hydrodynamics successfully characterize these flow observables, which makes the relativistic hydrodynamics the ``standard model" in heavy-ion collisions~\cite{Romatschke:2009im,Heinz:2013th,Gale:2013da,Yan:2017ivm,Florkowski:2017olj,Romatschke:2017ejr}.

Phenomenological analyses within hydrodynamic frameworks provide the most efficient, robust, and effective tool to extract the many-body QCD dynamics. For instance, as the macroscopic emergence of the interactions among quarks and gluons, transport coefficients in the QGP medium can be alluded from the comparisons between the hydrodynamic modeling and experimental data. By far, the specific shear viscosity, i.e., the ratio between the shear viscosity and the entropy density, $\eta/s$, has been constrained to values very close to a lower theoretical bound $\hbar/4\pi k_B$, suggesting QGP a strongly coupled medium~\cite{Shuryak:2008eq,Kovtun:2004de}. The specific bulk viscosity is extracted as well, leading to result with a temperature dependence \cite{Ryu:2015vwa}. In addition to the transport coefficients, properties of QCD are also hidden in the equation of state. These include the relations among the local energy density, the pressure, the entropy density and the speed of sound $c_s$, etc. Some undergoing attempts through the hydrodynamic modeling have achieved compatible results with the solutions of lattice QCD~\cite{Gardim:2019xjs,Gardim:2019brr}.

These efforts on studying transport coefficients and the equation of state based on hydrodynamics are essential towards a quantitative characterization of the QCD matter. Especially, a reliable hydrodynamic description of the system evolution is crucial for the search for a conjectured QCD critical point and its associated first-order phase transition between QGP and hadron gas at a finite baryon density~\cite{Stephanov:2004wx}. Searching for the QCD critical point is the focus of the current Beam Energy Scan program (BES) at RHIC. However, extensions of the hydrodynamic model to cases involving finite baryon densities are challenging, especially, considering the significant hydrodynamical fluctuations of baryon density associated with the QCD critical point. Difficulty stems not only from improving the model itself, such as including baryon charge in the equation of state, it also requires some fundamental progress in the theoretical formulation of hydrodynamics, so that stochastic hydrodynamical fluctuations can be taken into account systematically~\cite{Landau1987Fluid,Kovtun:2011np,Kapusta:2011gt,Young:2014pka,Akamatsu:2016llw, Singh:2018dpk, An:2019osr}, and novel hydrodynamic modes due to the effect of critical  slowing-down can be included~\cite{Stephanov:2017ghc,An:2019csj,Rajagopal:2019xwg,Du:2020bxp}.
When non-zero baryon density is involved, how the collective behavior of the QGP is changed, and correspondingly how the observed correlations of these generated hadrons are modified, needs to be answered in the hydrodynamic modeling.

The successes of hydrodynamics and its applications to heavy-ion collisions also bring us with many surprises. 
Although in large systems created in high-energy nucleus-nucleus collisions, various observables with respect to the collective flow have been found consistent with hydrodynamic modelings \cite{Schenke:2020mbo},
application of hydrodynamic modeling to small systems such as those created in proton-nucleus collisions~\cite{Khachatryan:2015waa,Khachatryan:2015lva, PHENIX:2018lia} is not straightforward, owing to the significant reduction of system size and strong expansion rate \cite{Niemi:2014wta, Kurkela:2019kip}. The ``unreasonable effectiveness"~\cite{Romatschke:2017ejr, Mantysaari:2017cni, Schenke:2019pmk, Schenke:2020mbo} in describing the collectivity in the small colliding systems has modified qualitatively the understanding of QCD system thermalization~\cite{Romatschke:2016hle}. 
The condition of the onset hydrodynamics (\emph{hydrodynamization}) is even relaxed. The traditionally recognized hydrodynamic and non-hydrodynamic modes, and propagation of these modes~\cite{Kurkela:2018wud,Kurkela:2018vqr}, have been generalized largely beyond local thermal equilibrium. Out-of-equilibrium hydrodynamics, a novel concept associated with the discovery of attrator solutions in various dynamical systems~\cite{Heller:2013fn}, has been proposed as a theoretical candidate to generalize the applicability of hydrodynamics. In the recent few years, a lot of progresses have been made to develop the theoretical formulation of the out-of-equilibrium hydrodynamics.

The successful phenomenological application of relativistic fluid dynamics in heavy-ion collisions and the continuous supports and challenges from the RHIC and LHC experiments have led to a vibrant program which unites research from traditionally separate disciplines such as string theory, computational physics, statistics, nuclear physics, and high-energy physics. Recent direct detection of gravitational waves from black holes and neutron star mergers \cite{Abbott:2016blz, Abbott:2016nmj, TheLIGOScientific:2017qsa} adds another interconnection with relativistic heavy-ion collisions at large baryon density.

This review will focus on these recent development in out-of-equilibrium hydrodynamics and highlight some of the current state-of-the-art phenomenological applications of hydrodynamic frameworks to describe the dynamics of relativistic heavy-ion collisions. 

In section~\ref{sec.II}, we review the theoretical formulation of out-of-equilibrium hydrodynamics at an introductory level. This is going to be presented first from the extension of second order viscous hydrodynamics to systems with large local gradients quantified by the Knudsen number in the Bjorken flow. Attractor solution from such a dynamical system emerges naturally, as a consequence of the existence of fixed points in both the free-streaming and hydrodynamic regimes. The relation between the attractor and the asymptotic hydrodynamic gradient expansion is addressed as well, in the context of the trans-series solution and resurgence properties in the theory of asymptotic series. An alternative approach from kinetic theory is discussed, in terms of a set of moments of the phase-space distribution function. These moments are coupled through their equations of motion. The lowest orders of the equation reduce to the familiar hydrodynamic equation of motion when gradients of system tend to vanish. Out-of-equilibrium effects can be accounted for by higher order moments, whose contribution to the system evolution out of equilibrium results in an effective correction of the transport coefficients.

Section~\ref{sec.III} covers the state-of-the-art applications of (3+1)D hydrodynamics + hadronic transport framework to heavy-ion collisions at intermediate and high collision energies. The experimental programs at $\sqrt{s} \sim \mathcal{O}(10)$\,GeV are extremely exciting to map out the phase structure of the QCD matter at finite net baryon densities. In the meantime, 3D hydrodynamic framework also opens a new dimension to study event-by-event fluctuations along the longitudinal direction. Because it is difficult to calculate the transport properties of the QGP from first principles, quantitative characterization of the QGP has been leading by phenomenological analysis. We will summarize the collective effort in constraining the specific shear and bulk viscosity over the past decades and highlight recent efforts towards accessing the baryon diffusion constant in QGP. While hydrodynamics becomes the standard theory to describe the large heavy-ion systems, smaller collisions in p+A and p+p collisions challenges the conventional picture of the validity region of hydrodynamics. We have started to see a connection between the leading development of out-of-equilibrium hydrodynamics formulation and strong collectivity in those small systems. Finally, we highlight some interdisciplinary connections between heavy-ion physics and nuclear structure physics as well as statistics and machine learning applications.

\section{Out-of-equilibrium hydrodynamics}\label{sec.II}

We start with a brief introduction on the fundamental concepts of viscous hydrodynamics that has been applied in the study of high energy heavy-ion collisions. As an essential ingredient of the hydrodynamic modeling, it should be emphasized that a truncation at the second order in gradients is generally considered in these viscous hydrodynamics formulation. This is to be distinguished from some recent development of out-of-equilibrium hydrodynamics which often involves gradients to infinite orders.

\subsection{Viscous hydrodynamics}

Hydrodynamics is a low energy effective theory, which describes the evolution of long wavelength modes in a dynamical system. These are slow modes, commonly known as hydrodynamic modes, obey a set of hydrodynamic equations of motion stemming from conservation laws. The conservation of energy and momentum, for instance, $\partial_\mu T^{\mu\nu} = 0$, plays a key role in 
determining the space-time evolution of the hydrodynamic fields: 
Local energy density $\e$, pressure $\P$ and fluid four-velocity $U^\mu$~\cite{Gale:2013da}.\footnote{
We take the normalization of four-velocity as $U^\mu U_\mu=1$, corresponding to the most negative metric convention:
$g_{\mu\nu}=(+,-,-,-)$. }
The general form of the energy-momentum tensor $T^{\mu\nu}$ is given in 
the corresponding constitutive relation as
\be
\label{eq:tmn0}
T^{\mu\nu}
=\e U^\mu U^\nu - (\P+\Pi) \Delta^{\mu\nu} + \pi^{\mu\nu}\,.
\ee
The projection tensor is defined as 
\be
\Delta^{\mu\nu}=g^{\mu\nu} - U^\mu U^\nu\,,
\ee
so that the spatial gradient can be formulated in a covariant form $\nabla^\mu = \Delta^{\mu\nu}\partial_\nu$. For late convenience, we also have the definition of co-moving time derivative $D=U^\mu \partial_\mu$, together with which a four-vector can be decomposed into a temporal component and a spatial component with respect to the fluid four-velocity $U^\mu$, respectively. Especially, the normal derivative can be separated as, $\partial_\mu = \nabla_\mu + U_\mu D$.

In addition to ideal hydrodynamics corresponding to a fluid system in local thermal equilibrium, 
\be
\label{eq:idealTmn}
T^{\mu\nu}_{\rm ideal}=\e U^\mu U^\nu - \P \Delta^{\mu\nu}\,,
\ee
there are viscous corrections in the energy-momentum tensor $T^{\mu\nu}$ to capture deviations of the fluid system from local thermal equilibrium. In the framework of viscous hydrodynamics, these corrections are formulated in terms of an expansion over spatial gradients of the hydrodynamical fields. More precisely, this expansion is characterized by the Knudsen number, Kn, which is essentially the dimensionless ratio between a microscopic length scale and a macroscopic length scale. For the bulk pressure $\Pi$ and the shear stress tensor $\pi^{\mu\nu}$, one has up to the first order in gradient, the Navier-Stokes hydrodynamics,
\begin{align}
\label{eq:NSTmn}
\pi^{\mu\nu} &= 2\eta ^\bra \nabla^\mu U^{\nu\ket} + O(\nabla^2)\,,\cr
\Pi & =- \zeta \nabla\cdot U
+ O(\nabla^2)\,,
\end{align}
where $\eta$ and $\zeta$ are the shear and bulk viscosities, respectively. These are transport coefficients
determined by interactions among fluid constituents, reflecting the dynamic nature of underlying theories. For \Eq{eq:NSTmn}, the Knudsen number can be read off roughly as \cite{Niemi:2014wta},
\be
\Kn \sim \frac{|\eta ^\bra \nabla^\mu U^{\nu\ket}|}{\P} \quad \mbox{ or }\quad
\frac{|\zeta \nabla\cdot U|}{\P}\,.
\ee
In \Eq{eq:NSTmn} and in what follows, the brackets around tensor indices indicate a symmetric, transverse and traceless projection of a tensor, \ie,
\be
{}^\bra A^{\mu\nu\ket}=\Delta^{\mu\nu\alpha\beta}A_{\alpha\beta}\,,
\ee
where
\be
\Delta^{\mu\nu\alpha\beta}\equiv
\frac{1}{2}\left[\Delta^{\mu\alpha}\Delta^{\nu\beta}+\Delta^{\mu\beta}\Delta^{\nu\alpha}\right]-\frac{1}{3}\Delta^{\mu\nu}\Delta^{\alpha\beta}\,.
\ee

In spirit of the hydrodynamic gradient expansion, \Eq{eq:NSTmn} 
can be systematically extended to higher orders. In particular, considering the fact that the resulted equations of motion from Navier-Stokes hydrodynamics are acasual, the extension to elevating the dissipative currents to dynamical degrees of freedom is necessary. \footnote{ Causality and stability condition can be achieved in first order viscous hydrodynamics as well, but within a frame other than the choice by Landau-Lifshitz or Eckart~\cite{Kovtun:2019hdm}.)
}
For practical simulations, acasual modes can be remedied by using the Israel-Stewart formulation~\cite{Israel:1979wp}, with second order gradient terms included. These terms relax to its Navier-Stokes form, with the relaxation effect specified by correspondingly the shear and bulk relaxation time: $\tau_\pi$ and $\tau_\Pi$. With respect to conformal symmetry the second order shear stress tensor is completely determined by the BRSSS hydrodynamics~\cite{Baier:2007ix},
\begin{align}
\label{eq:brsss}
\pi^{\mu\nu} =&\; \eta^{\bra}\nabla^\mu U^{\nu\ket} - \tau_\pi\left[
{^\bra}D\pi^{\mu\nu\ket} + \frac{4}{3}\pi^{\mu\nu}\nabla\cdot U
\right] \cr 
&-\frac{\lambda_1}{\eta^2}\pi^{\bra\mu}_{\quad\alpha}\pi^{\nu\ket\alpha}
-\frac{\lambda_2}{\eta}\pi^{\bra\mu}_{\quad\alpha}\Omega^{\nu\ket\alpha}
- \lambda_3 \Omega^{\bra\mu}_{\quad\alpha}\Omega^{\nu\ket\alpha}\,,\cr
\end{align}
where in addition to the shear relaxation time $\tau_\pi$, $\lambda_1$, $\lambda_2$ and $\lambda_3$ are independent second order transport coefficients. For conformal fluids, these transport coefficients are known~\cite{Kovtun:2004de,Baier:2007ix} and can be parameterized as~\cite{Heller:2015dha,Basar:2015ava}
\be
\eta = C_\eta s,\quad
\tau_\pi = \frac{C_\tau C_\eta}{T},\quad
\lambda_1 = C_{\lambda_1} \frac{s}{T}\,,
\ee
where the local entropy density $s\propto T^3$. For consistency and considering a weakly coupled system, in the current review we shall take the evaluations from kinetic theory for a conformal system, that~\cite{Kovtun:2004de,Dusling:2009df}\footnote{
These transport coefficients have different evaluations for a strongly coupled system. From the $\mathcal{N}=4$ supersymmetric Yang-Mills theory, they are~\cite{Baier:2007ix}
\be
C_\tau = \frac{2-\log 2}{2\pi}\,,\quad
C_\lambda = \frac{1}{2\pi}\,.
\ee
},
\be
C_\eta = \frac{1}{4\pi}\,,\quad
C_\tau = 5\,,\quad
C_{\lambda_1} = \frac{5}{7}C_\eta C_\tau\,.
\ee

There exists other variant form of second order viscous hydrodynamics 
in addition to \Eq{eq:brsss}, when conformal symmetry is not guaranteed~\cite{Denicol:2012cn}. 
Note that \Eq{eq:brsss} is consistent
with the M\"uller-Israel-Stewart theory~\cite{Israel:1979wp}, which does relaxes to \Eq{eq:NSTmn} when relaxation time
$\tau_\pi \to 0$. Note also that there are more tensor structures aroused in the second order terms, such
as the vorticity tensor $\Omega^{\mu\nu}$, $\Omega^{\mu\nu} = \nabla^\mu U^\nu - \nabla^\nu U^\mu$, while for the Navier-Stoke hydrodynamics only one term is involved. 

Extension to even higher orders has been considered in literature (cf.~\cite{El:2009vj,Jaiswal:2013vta}), with more and more tensor
structures introduced together with correspondingly new transport coefficients. As a consequence of the 
increasing number of tensor structures, it is expected 
at n-th order, the number of new transport coefficients, or to say, the number of new tensor structures scales as
$n!$. This factorial increase essentially affects the convergence properties of hydrodynamic gradient expansion, that 
hydrodynamic gradient expansion is rather asymptotic than convergent~\cite{1963PhFl....6..147G,PhysRevLett.110.211602,denicol2016divergence}.\footnote{Convergence of hydro gradient expansion depends also on the detailed identification of the expansion parameter. For instance, the dispersion relation consisting of perturbations around equilibrium gives rise to a series expansion in terms of the wave-number, which is convergent[cf. Ref~\cite{Grozdanov:2019kge}]. 
On the other hand, for series expansion in real space over spatial gradients, convergence property may depend on initial condition~\cite{Heller:2020uuy}. }
In addition to the shear channel, the 
asymptotic property of the hydro gradient expansion exists in the bulk and diffusion channels as well. 
In principle, applicability of the classical
framework of hydrodynamics relies on the analysis of gradient expansion. 

\begin{figure*}
    \centering
    \includegraphics[width=0.9\linewidth]{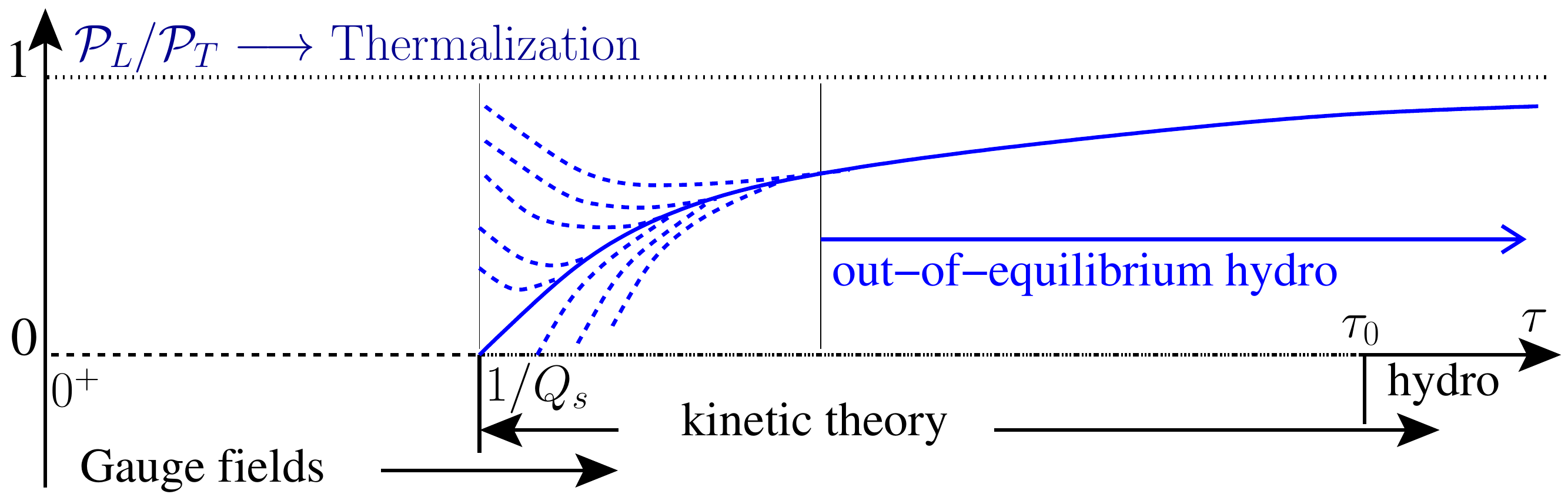}
    \caption{Schematic illustration of the early-time stages of the system evolution in high-energy heavy-ion collisions, describable respectively starting from initial time of heavy-ion collisions by classical gauge theory for the gluon field, then around $\tau\sim 1/Q_s$ by kinetic theory for quarks and gluons, and by fluid dynamics for the hydrodynamic variables such as pressure, energy density after hydrodynamization. The blues lines are pressure ratio between the longitudinal and transverse pressures, $\P_L/\P_T$, characterizing how far the system is away from local thermal equilibrium, \ie, $\P_L/\P_T=1$.}
    \label{fig:ic_evolve}
\end{figure*}

In a similar way, the charge conservation gives, $\partial_\mu J^\mu = 0$, where the conserved current of hydrodynamics receives dissipative corrections as well,
\be
\label{eq:jmn0}
J^\mu \equiv J^\mu_{\rm ideal} + I^\mu = n U^\mu + I^\mu \,,
\ee
with 
\be
\label{eq:NSJm}
I^\mu = \sigma T \nabla^\mu \left(\frac{\mu}{T}\right) + O(\nabla^2)\,,
\ee
and $\sigma$ being the corresponding conductivity of the conserved charge. For heavy-ion collisions, the net baryon
number, which is related to the QCD critical behavior, is commonly considered in a hydrodynamic analysis. As in the shear channel, acasual modes can be avoided by extending the constitute relation in \Eq{eq:NSJm} to the Cattaneo equation with a finite relaxation time $\tau_Q$ \cite{Kapusta:2017hfi},
\be
I^\mu = \sigma T \nabla^\mu (1 + \tau_Q D )^{-1} \left(\frac{\mu}{T}\right)  \,.
\ee

For most theoretical analyses carried out with respect to the QGP in high energy nucleus-nucleus collisions, with up to second order viscous corrections, the aforementioned equations provide the essential ingredient in a successful phenomenological model which suffices to capture the system evolution. Together with an equations of state provided by lattice QCD simulations, \eg, $\P = \P(e)$, numerical solutions to the hydrodynamic modeling gives rise to space-time evolution of the hydrodynamic fields, which eventually reaches freeze-out and yields the observed particles in experiments.
Some more details on the phenomenological modeling will be given later in \Sect{sec.III}.

\subsection{Hydrodynamization and out-of-equilibrium fluid dynamics} 

Hydrodynamic modeling has been successfully applied to small colliding systems as well. For the high multiplicity events of proton-lead collisions~\cite{CMS:2012qk,Khachatryan:2015waa,Sirunyan:2019pbr,Acharya:2019vdf}, $^3$He-gold collisions, deuteron-gold collisions~\cite{Adare:2015ctn,Adare:2017wlc,Lacey:2020ime}, and even proton-proton collisions~\cite{Khachatryan:2015lva}, the observed multi-particle correlations are found compatible with a hydrodynamic prediction~\cite{Nagle:2013lja,Habich:2015rtj}. Comparing to nucleus-nucleus collisions, in these systems, the created QGP fireball is expected to be small and short-lived, which in turn gives large spatial gradients. The question of why hydrodynamics being `unreasonably' successful in small colliding systems is one of the recent focuses in the heavy-ion community, which has motivated extensive theoretical development of the out-of-equilibrium hydrodynamics~\cite{Heller:2015dha,Heller:2018qvh,Denicol:2018wdp,Florkowski:2017olj,Strickland:2017kux,Behtash:2017wqg,Romatschke:2017acs,Romatschke:2017vte,Kurkela:2019set,Denicol:2019lio,Chattopadhyay:2019jqj,Strickland:2018ayk,Brewer:2019oha,Blaizot:2017ucy,Behtash:2019txb,Dash:2020zqx,Chattopadhyay:2019jqj,Behtash:2019qtk,Heller:2020anv}.

When applying to realistic simulations of heavy-ion collisions, the framework of viscous hydrodynamics assumes a valid
truncation of the gradient expansion at second order in the gradients. The validity of the truncation requires that
the QGP system is locally close to thermal equilibrium such that in the equation of motion $O(\Kn^3)\ll O(\Kn^2)$, in this way gradient corrections of order higher than or equal to $O(\Kn^3)$ can be safely neglected. 
In heavy-ion collisions, this is by assumption satisfied at a time scale
$\tau_0$ when the QGP created from heavy-ion 
collisions approaches local thermal equilibrium (\emph{thermalization}), or when the system evolution starts to be captured by the 
second order fluid dynamics (\emph{hydrodynamization}). It should be emphasized that hydrodynamization is a much relaxed
condition than thermalization, which does not require isotropization between the longitudinal and transverse pressures, and the finite pressure difference $\P_L-\P_T$, is accounted for by dissipative effects. 

Comparing to the lifetime $\tau_f$ of QGP in heavy-ion collisions, 
it is obvious that only when $\tau_0\ll\tau_f$, can the fluid dynamics description of the system evolution be reliable. 
For a strongly coupled QGP medium, theoretical analysis based on AdS/CFT estimates that hydrodynamization scales inversely to temperature, 
$\tau_0\sim 1/T$~\cite{Chesler:2016ceu}. On the other hand, if the 
QGP system is weakly coupled, that the fundamental interactions are scatterings among quarks and gluons characterized by perturbative QCD, the onset of hydrodynamics is related to the strong coupling constant $\alpha_s$, and saturation scale $Q_s$, that $\tau_0\sim \alpha_s^x Q_s^{-1}$~\cite{Baier:2000sb} with the exponent $x$ being a constant negative number. Note that, however, given the actual values of strong coupling constant and $Q_s$ in heavy-ion collisions, the expected $\tau_0$ in a realistic QGP systems could be rather large~\cite{Baier:2000sb}. We shall concentrate on the weakly coupled system in the current discussion.
 
Despite all the estimates of $\tau_0$ in various theories, how the created system in heavy-ion collisions evolves towards fluids is itself an outstanding question.
An schematic illustration of the early stages of system evolution in heavy-ion collisions, considering more realistic situations that are compatible with the QCD dynamics, is shown in \Fig{fig:ic_evolve}. The created medium is believed to experience at first the stage of classical gluon field. This is the color glass condensate (CGC) picture~\cite{Gelis:2010nm}, 
in which the system evolution is dominated by saturated gluon field. Longitudinal and transverse pressure in the gluon field is highly anisotropic~\cite{Gelis:2013rba}. As the system expands the density of gluon field decreases, around $\tau\sim 1/Q_s$, a kinetic theory description becomes available for gluons. In the kinetic theory description, isotropization is eventually achieved via scatterings among quarks and gluons, against the effect of longitudinal expansion. 
In \Fig{fig:ic_evolve}, isotropization is characterized in by the the ratio between longitudinal and transverse pressures (blue lines), $\P_L/\P_T$. The blue dashed lines and the solid line respectively correspond to various initializations due to quantum fluctuations around $1/Q_s$, leading to different pressure anisotropies $\P_L/\P_T$ when kinetic theory description starts. At late times, regardless of arbitrary initial conditions, the evolution of pressure anisotropy becomes universal. At around $\tau_0$, the second order viscous hydrodynamics starts to dominate, while $\P_L/\P_T$ approaches unity, i.e., close to local thermal equilibrium.

The existence of such an universal evolution in \Fig{fig:ic_evolve} has been proved in various theoretical analyses. It implies uniquely the value of $\tau_0$, irrespective of initial conditions. Moreover, it provides a novel and extended description of the system evolution that applies to the system evolution out-of-equilibrium. This is the fundamental idea of out-of-equilibrium hydrodynamics. This universal evolution, which is dubbed ``attractor", can be shown beyond the characterization of hydrodynamics including infinite orders in the gradient expansion. 
So far, the studies based on attractor solution has been found theoretically feasible in some highly symmetric expanding systems, 
such as the Bjorken flow~\cite{Heller:2013fn,Bjorken:1982qr,Heller:2015dha,Basar:2015ava,Denicol:2018wdp,Denicol:2014ywa,Strickland:2017kux,Jaiswal:2019cju,Romatschke:2017vte,Kurkela:2019kip,Chattopadhyay:2019jqj,Blaizot:2017lht,Heller:2020anv} and Gubser flow~\cite{Gubser:2010ui,Denicol:2018pak,Dash:2020zqx,Behtash:2019qtk,Behtash:2017wqg}, and numerically in less symmetric systems~\cite{Romatschke:2017acs}.
It has been solved with respect to the equation of motion from hydrodynamics, as well as kinetic theory. From either perspective, we shall present a pedestrian introduction to the derivation of the attractor solutions.

\subsubsection{From fluid dynamics}

As an example, we first present the analysis with respect to conformal fluids and Bjorken symmetry. Bjorken symmetry is a good approximation for the system created in high energy heavy-ion collisions in its very early stages~\cite{Bjorken:1982qr}. It describes the longitudinal boost-invariant expansion of the medium along the beam axis (which we identify as $z$), while transverse expansions along $x$ and $y$ are ignored. The boost invariant symmetry is motivated by the observations in high energy collisions at the mid-rapidity region~\cite{Bjorken:1982qr}, while considering the fact that longitudinal expansion dominates at the stage shortly after collisions allows one to ignore transverse expansions. 

Instead of using the usual Minkowski space-time coordinates $t$ and $z$, with the new Milne coordinates, 
\be
\tau=\sqrt{t^2-z^2}\,,\qquad
\xi =\tanh^{-1}(t/z)\,,
\ee
the Bjorken flow is simplified in the way that hydrodynamic fields can be written 
independent of the space-time rapidity $\xi$. In the Milne coordinate system,
the fluid four-velocity under Bjorken symmetry is fully determined as,
\be
U^\tau=1\,,\qquad
U^x=U^y=U^\xi=0\,.
\ee
Therefore, taking into account the fact that the metric is rewritten as 
$ds^2=g_{\mu\nu} dx^\mu dx^\nu=d \tau^2-\tau^2 d\xi^2 - d x^2_\perp$, gradients of 
the fluid fields are reduced and all related to the proper time $\tau$, e.g.,
\be
\label{eq:bhydro}
\nabla\cdot U = -\frac{3}{4}\sigma^\xi_{\,\,\xi} = \frac{1}{\tau}\,.
\ee
This relation implies one unique expression of the Knudsen number for the Bjorken symmetry,
that is, $\Kn^{-1}\sim\tau/\tau_\pi$. 

The equation of motion of hydrodynamics $\partial_\mu T^{\mu\nu}=0$ becomes
\be
\label{eq:eom_ehydro}
\partial_\tau\epsilon =-\frac{4}{3}\frac{ \epsilon}{\tau} + \frac{\pi^\xi_{\,\,\xi}}{\tau}\,.
\ee
Without dissipation, the above equation has an analytic solution for the energy density, the ideal hydrodynamic evolution, $\epsilon\sim \tau^{-4/3}$. Note that the exponent $-4/3$ is the characteristic decay rate of energy density in the ideal and conformal fluid experiencing Bjorken expansion. With dissipative corrections, the expected full solution of energy density consists of gradient corrections,
\be
\label{eq:energyI}
\epsilon \sim \tau^{-4/3}\left(1+\sum_n O\left(\frac{\tau_\pi}{\tau}\right)^n\right)\,.
\ee

One may check that in the conformal fluid with respect to Bjorken symmetry, only $\pi^\xi_{\,\,\xi}=-2\pi^x_{\,\,x}=-2\pi^y_{\,\,y}$ are the non-zero components of the shear stress tensor. Similarly, the vorticity tensor $\Omega^{\mu\nu}$ vanishes. Defining $\pi=\pi^\xi_{\,\xi}$, and given all the information in Bjorken flow, the constitutive relation for $\pi$ follows from the BRSSS theory, \Eq{eq:brsss},
\be
\label{eq:Phi_brsss}
\pi=-\frac{4}{3}\frac{\eta}{\tau} - \tau_\pi \left[\frac{\rmd \pi}{\rmd \tau} + 
\frac{4}{3}\frac{\pi}{\tau}\right] + \frac{\lambda_1}{2\eta^2} \pi^2\,,
\ee
which is a nonlinear first order differential equation. Without the constraint of conformal symmetry, 
the equation of motion is not unique. For instance, it can be also formulated according to the DNMR approach as~\cite{Denicol:2012cn}
\be
\label{eq:Phi_DNMR}
\pi=-\frac{4}{3}\frac{\eta}{\tau} - \tau_\pi \left[\frac{\rmd \pi}{\rmd \tau} + 
\beta_\pi \frac{\pi}{\tau}\right] - \frac{\chi \tau_\pi^2}{\eta} \frac{\pi^2}{\tau}\,,
\ee
where $\beta_\pi$ and $\chi$ are transport coefficients in this alternate formulation. Note that $\beta_{\pi}$ and $\chi$ are related to the second order transport coefficients appear in the BRSSS hydrodynamics (cf. \Eq{eq:betalambda}). It is worth mentioning that, for some evaluations of the transport coefficients, e.g., constant $\tau_\pi$, the coupled equations of motion in \Eq{eq:Phi_brsss} and \Eq{eq:Phi_DNMR} can be solved analytically~\cite{Denicol:2017lxn,Jaiswal:2019cju}. 

To proceed, one needs to solve the coupled equations, \Eqs{eq:eom_ehydro} and 
(\ref{eq:Phi_brsss})
and to construct the hydro gradient expansion accordingly. A convenient way to do so is to introduce
\be
\label{eq:g0def}
g(w) \equiv \frac{\rmd \,\ln \epsilon}{\rmd \,\ln \tau}=-1-\frac{P_L}{\e}\,,
\ee
as a function of the inverse Knudsen number
\be
w \equiv \tau/\tau_\pi\,. 
\ee
Apparently, $g(w)$ characterizes the decay rate of energy density, and is related to the pressure ratio.
Although in the conformal case, $\rmd w/\rmd \tau>0$ is satisfied, $w$ is more significant than purely a time
scale. A small $w$ could indicate early time of the system evolution, 
but it can be also interpreted as the system being far away from local equilibrium.
On the other hand, a large $w$ could indicate late time, but it also corresponds to systems close to local equilibrium. 

\begin{figure}
    \centering
    \includegraphics[width=1.0\linewidth]{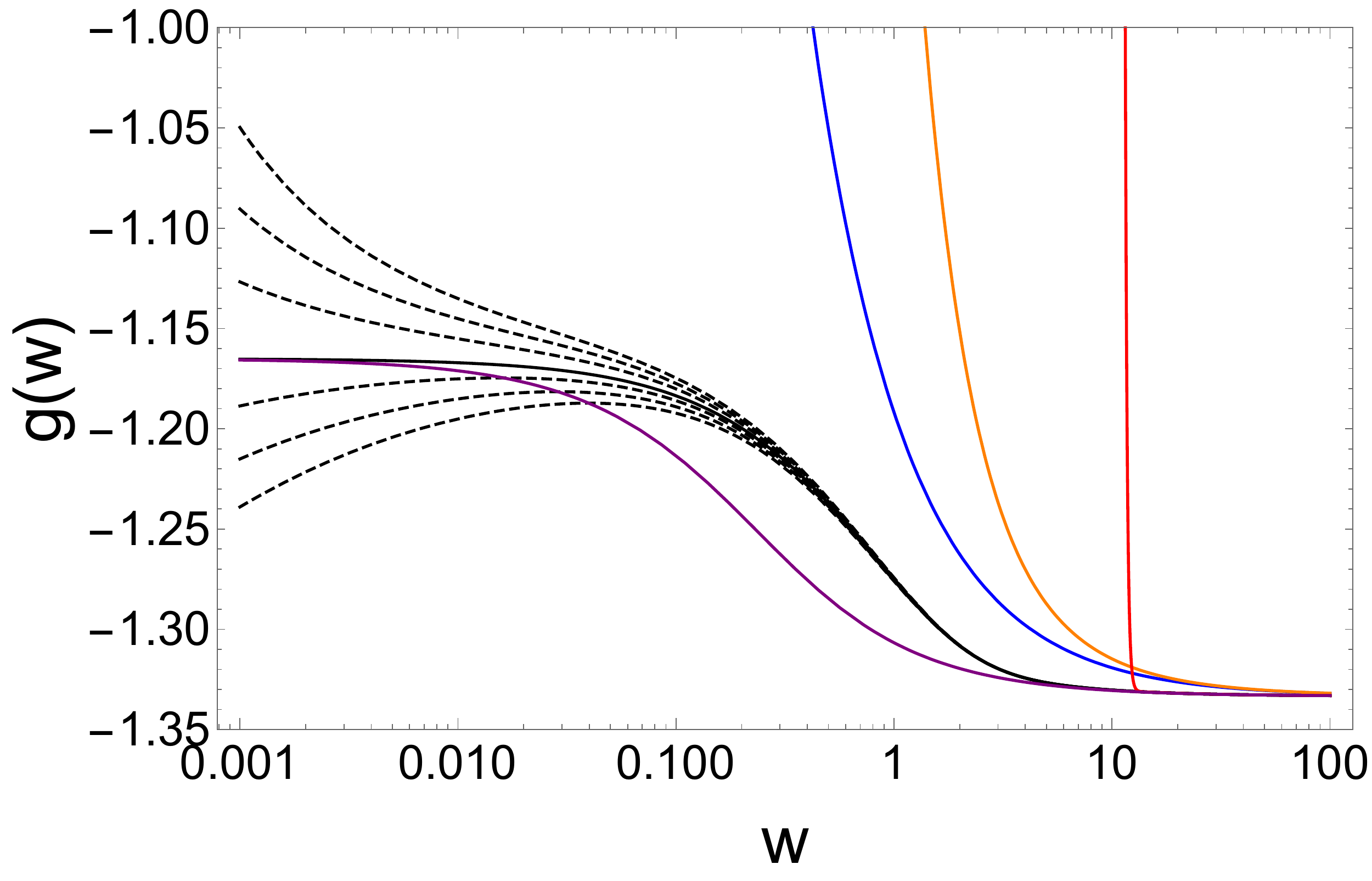}
    \caption{Numerical solution of $g(w)$ to the conformal second order fluid dynamics with respect to various initial conditions (dashed lines). The solid black line corresponds to the attractor solution, starting from the free-streaming stable fixed point $g_+$ of the evolving system when $w\to 0^+$. Solutions from random initial conditions (dashed lines) converge towards the attractor solution when $w\sim 1$. A leading order slow-roll approximation of the attractor
    solution is given as the solid purple line~\Eq{eq:slowroll}. Results with corrections from first order, second order and up to 50th order in viscous hydrodynamics are plotted as the solid blue, orange and red lines, respectively.}
    \label{fig:attractor_BRSSS}
\end{figure} 

After some algebra, rewriting the coupled equations in terms of $g(w)$, one realizes a nonlinear first order
differential equation,
\begin{align}
\label{eq:gBRSSS}
 w &\frac{{\rmd }g}{{\rmd} w}\left(1+\frac{g}{4}\right)+\left(g+\frac{4}{3}\right)^2\left[1+\frac{3w}{8}\frac{C_{\lambda}}{C_\eta}\right] \nonumber\\
 &+w\left(g+\frac{4}{3}\right)-\frac{16}{9}\frac{C_\eta}{C_\tau}=0\,,
\end{align}
where the transport coefficients have been substituted by the parameterization constants.

With initial conditions of the function $g(w)$ given, one can solve \Eq{eq:gBRSSS}. Numerical solutions are presented in \Fig{fig:attractor_BRSSS}, together with the results obtained from viscous hydrodynamic solution with viscous corrections up to the first order (solid blue line), the second order (solid orange line) and 50th order terms, respectively. In comparison with the full solutions, it is apparent that hydrodynamics with truncated viscous corrections is valid only when $w\gg1$, as anticipated. Moreover, including more viscous corrections does not help to improve the solution towards out of equilibrium. This is a direct consequence that hydro gradient expansion is not convergent, as we shall detail later. 

As shown in \Fig{fig:attractor_BRSSS}, full numerical solutions starting from various initial conditions merge to one single curve, indicated as the solid black line, around $w\lesssim1$. This solid black line is known as the attractor solution, which in a dynamical system is often referred to solutions collected in the phase space irrespective of initial conditions~\cite{Heller:2020anv}. This attractor solution in hydrodynamics has been first noticed in the context of Bjorken flow~\cite{Heller:2013fn,Heller:2015dha}, and later in the Gubser flow (cf. Ref~\cite{Denicol:2017lxn}). It is also realized in the solution of kinetic equation for weakly coupled media and strongly coupled media using AdS/CFT~\cite{Romatschke:2017acs,Kurkela:2019set}. 
The attractor solution offers a valid and universal description of the system evolution, even if $w\lesssim1$. It therefore largely extends the applicability of hydrodynamics to out-of-equilibrium systems. 

Numerically, attractor solution can be solved with respect to one special initial condition, corresponding to the free-streaming stable fixed point. Analytically, the emergence of an attractor solution in the nonlinear differential equation can be understood either in
terms of the evolution of (pseudo) fixed points, or the Borel resum of the hydrodynamic gradients. 

\subparagraph{Fixed point analysis. }
\Eq{eq:gBRSSS} has singularities at $w=0$ and infinity, between which the solution is expected to be analytic.
Hence it is worth examining the two extremes: far-from-equilibrium extreme with $w\to 0^+$ and close-to-local-equilibrium extreme with $w\to +\infty$. In particular, one should concentrate on the stable fixed points in these extremes that govern the system evolution.

Because $w\to 0^+$ is equivalent to setting $\tau_\pi\to \infty$, which corresponds to infinitely weak interactions 
among fluid constituents, the far-form-equilibrium extreme can be also interpreted as system evolution determined entirely by expansion. This is known as  free streaming. 
In the limit of small $w$, the nonlinear differential equation reduces to an algebraic equation,
\be
\left(g+\frac{4}{3}\right)^2-\frac{16}{9}\frac{C_\eta}{C_\tau}=0\,,
\ee
with its two solutions given as ($g_-<g_+$)
\be
\label{eq:fp0}
g_\pm = -\frac{4}{3}\left( 1\mp \sqrt{\frac{C_\eta}{C_\tau}}\right)\,.
\ee
These are the two fixed points of free streaming, and $g_+$ corresponds a stable point while
$g_-$ leads a unstable fixed point. That is to say, had the system been initialized at $g_+$, it will stay at 
$g_+$ for a purely expanding system. In the pure free streaming systems, perturbations around the 
stable fixed point decay with time, and the decay rate scales as a power law~\cite{Kurkela:2019kip,Chattopadhyay:2019jqj},
\be
\delta g(w) \sim \left(\frac{\tau}{\tau_{\rm ini}}\right)^{-(g_+-g_-)}\,.
\ee
This power law decay qualitatively explains the observed pattern of $g(w)$ evolution at early times in \Fig{fig:attractor_BRSSS}.

In the opposite limit that $w\to +\infty$, Knudsen number is small enough that the system approaches the hydrodynamic regime. 
Given \Eq{eq:energyI} and the definition of $g(w)$,  
it is not difficult to notice that in the limit of $w\to+ \infty$, 
\be
\label{eq:fp1}
g(w)\to g_{\rm hyd}=-\frac{4}{3}\,,
\ee
which is the energy density decay rate of ideal hydrodynamics. This is the hydrodynamic fixed point,
to be reached as long as hydrodynamization being realized in a system.

\begin{figure}
    \centering
    \includegraphics[width=1.0\linewidth]{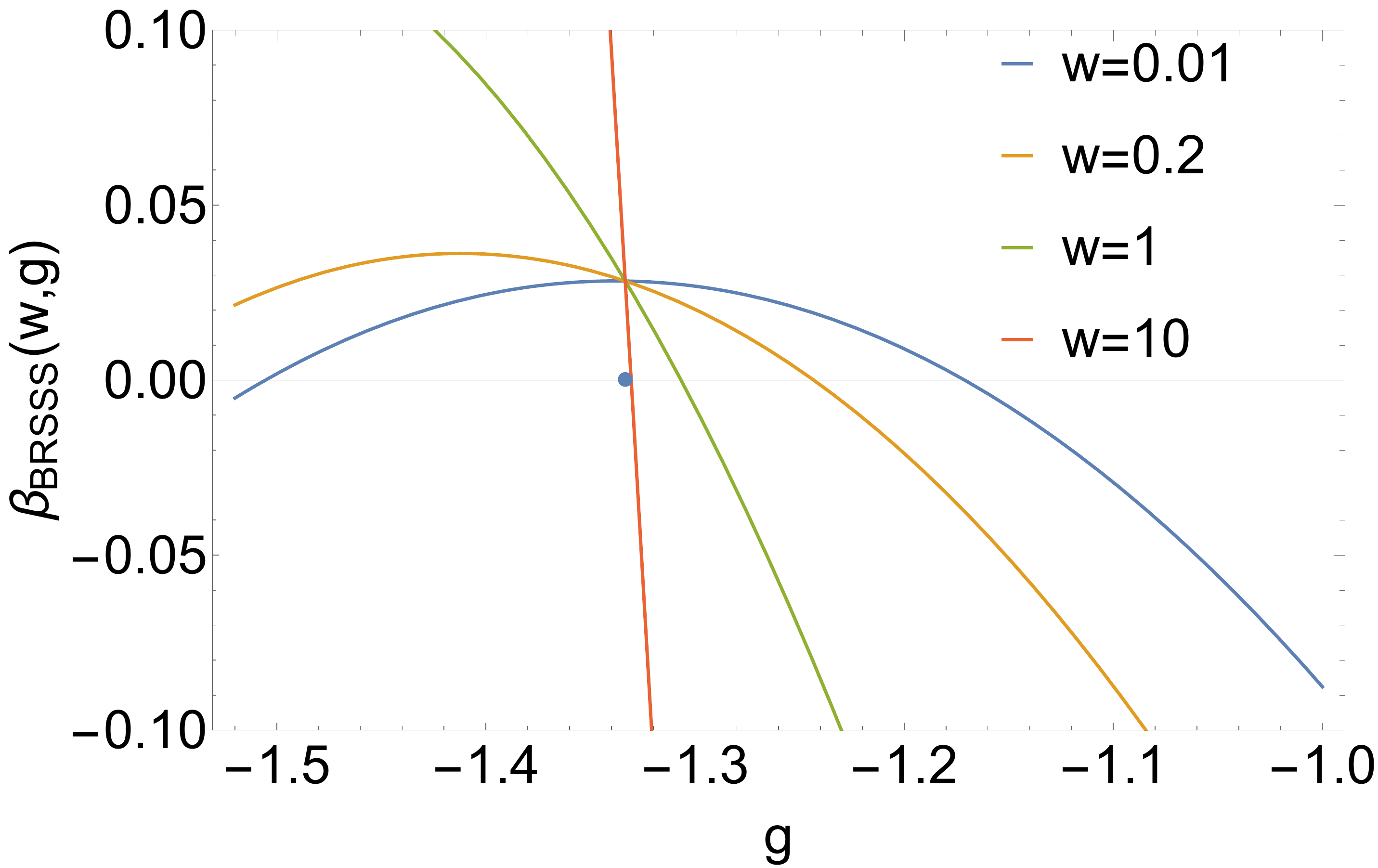}
    \caption{Numerical evaluation of the beta function \Eq{eq:betaBRSSS} for different values of $w$. Crossing with the x-axis stands for the (pseudo) fixed point at the corresponding $w$. The blue dot indicates the location of the hydrodynamic fixed point, corresponding to $w\to \infty$. A similar figure can be found in \cite{Blaizot:2019scw}. }
    \label{fig:betaBRSSS}
\end{figure}

In fact, a convenient way to reveal the properties of these fixed points, and to understand how these fixed points emerge from 
\Eq{eq:gBRSSS} in both extremes, is to define effectively a beta function. By looking at the root to $dg/dw = 0$, one defines effectively
\begin{align}
\label{eq:betaBRSSS}
\beta_{\rm BRSSS}(w, g) \equiv & \left(g+\frac{4}{3}\right)^2\left[1+\frac{3w}{8}\frac{C_{\lambda}}{C_\eta}\right] 
\nonumber\\
 & +w\left(g+\frac{4}{3}\right)-\frac{16}{9}\frac{C_\eta}{C_\tau}\,.
\end{align}
The root of Eq.~(\ref{eq:betaBRSSS}) encodes the information of fixed points in \Eq{eq:fp0} and in \Eq{eq:fp1}. 
Figure~\ref{fig:betaBRSSS} shows an illustration of the beta function for various values of $w$, from a small value ($w=0.01$) in the vicinity of free streaming, to a large value ($w=10$) in the hydrodynamic regime. The root of the beta function
is seen as the solved line crossing x-axis, and the stable fixed point corresponds to the crossing with a negative slope. Indeed, as $w$ increases from 
$w=0.01$, where the crossing gives rise to a stable
fixed point around $g_+\approx -1.165$ (and an unstable fixed point around $g_-\approx -1.502$ at which $dg/dw>0$),
the stable fixed point moves smoothly towards the hydro fixed point at $g_{\rm hyd}=-4/3$. Note that the hydro fixed point is super stable, since it is related to the crossing with an infinite negative slope. Accordingly, the unstable fixed point evolves, from the free streaming to a super unstable fixed in the hydro regime, around $g=-\frac{4}{3}\left(1+\frac{2 C_\eta}{C_{\lambda_1}}\right)=-2.08$, which again, is a root to the $\beta_{\rm BRSSS}=0$ in the limit of $w\to +\infty$.

Approximately, one may consider the accumulation of all these solved stable fixed points from the beta function form an analytical representation for the attractor. This is shown in \Fig{fig:attractor_BRSSS} as the purple solid line. Indeed, this approximation captures correctly the system evolution in both extremes,
but deviates when $w\sim 1$, even though this deviation is relatively small. This approximation procedure coincides with the leading order approximation using the slow-roll expansion~\cite{Liddle:1994dx,Heller:2015dha,Romatschke:2017vte,PhysRevD.97.056021}, that one neglects the derivative in \Eq{eq:gBRSSS} for the lowest order estimate: $d g/d w\sim0$,
\begin{align}
\label{eq:slowroll}
g_{\rm slow-roll}(w) = 
&-\frac{4}{3}+\frac{1}{2+\frac{3w}{4}\frac{C_\lambda}{C_\eta}}\Big[-w \cr
&+\sqrt{
w^2+\frac{16}{9}\frac{C_\eta}{C_\tau}\left(4+\frac{3w}{2}\frac{C_\lambda}{C_\eta}\right)}\Big]
\end{align} 
It is also compatible with the adiabatic evolution of a ground-state eigenmode (slowest mode), determined by the linear system of the original coupled hydrodynamic equation of motion~\cite{Brewer:2019oha}.

\subparagraph{Hydro gradient expansion, trans-series and resurgence.}
Hydrodynamic gradient expansion is a power series in term of the Knudsen number, i.e., $1/w$,
\begin{equation}
\label{eq:BRSSS_ge}
g_{{\rm hydro}(w)} \equiv
\sum_{n=0}^{\infty}\frac{f_n}{w^n}\,.
\end{equation}
Substituting the expansion into \Eq{eq:gBRSSS}, one finds the recursion relation for the expansion coefficients\footnote{
Note that this recursion relation differs from the one in~\cite{Basar:2015ava}, by a rescaling $w\to C_\tau w$ and $f_n\to 1-f_n/4$.
},
\begin{align}
\label{eq:fns}
&-\left(n-\frac{8}{3}\right) f_n + \sum_{k=0}^n\left(1 - \frac{k}{4} \right) f_k f_{n-k} 
+\left(1+\frac{C_\lambda}{C_\eta}\right)f_{n+1}\nonumber\\
&+\frac{3}{8} \frac{C_\lambda}{C_\eta}\sum_{k=0}^{n+1} f_k f_{n+1-k}+\left(1-\frac{C_\eta}{C_\tau}\right)\frac{16}{9} \delta_{n,0}\nonumber\\
&+\frac{2}{3}\left(2+\frac{C_\lambda}{C_\eta}\right) \delta_{n,-1}=0\,.
\end{align}
The leading order solutions of the above equation are 
\be
f_0 = -\frac{4}{3}\quad{\rm and}\quad
 -\frac{4}{3}\left(1+\frac{2 C_\eta}{C_{\lambda}}\right)\,,
\ee 
in agreement with those fixed points in the hydrodynamic regime found 
earlier from the root of the beta function, as anticipated.
The super-unstable fixed point is nonphysical, and it is not expected in realistic solutions. Starting from the hydrodynamic fixed point $f_0=-4/3$, using \Eq{eq:fns} iteratively, one is able to obtain the expansion coefficients to arbitrary orders. For instance, one finds, $f_1=\frac{16}{9} \frac{C_\eta}{C_\tau}$. In fact, it can be shown that magnitudes of these coefficients have a factorial growth, $f_n\sim n!$. This factorial growth can be recognized by noticing the ratio 
\be
\frac{f_{n+1}}{f_n}\to S^{-1}(n+\beta) + O\left(\frac{1}{n}\right)\,, 
\ee
for large $n$ in \Eq{eq:fns}. The parameters $S$ and $\beta$ are real constants. With respect to \Eq{eq:gBRSSS}, they are determined as
\be
\label{eq:sbeta}
S = \frac{3}{2}\,,\qquad
\beta^{-1} = -\frac{2C_\eta}{C_\tau}\,.
\ee
Therefore, the leading order contribution to the coefficients at very large $n$ is
$f_n\sim \Gamma(n+\beta)/S^{n+\beta}\sim n!$.
The factorial growth of the expansion coefficients results in the well-know statement that hydrodynamic gradient expansion is rather asymptotic than convergent, which has a vanishing the radius of convergence. Asymptotic series are commonly seen in physics, such as the perturbative expansion in quantum field theory~\cite{PhysRev.85.631,PhysRevD.87.025015,Blaizot:2003iq} and WKB approximation in quantum mechanics~\cite{ZINNJUSTIN2004197,ZINNJUSTIN2004269}.

\begin{figure}
    \centering
    \includegraphics[width=1.0\linewidth]{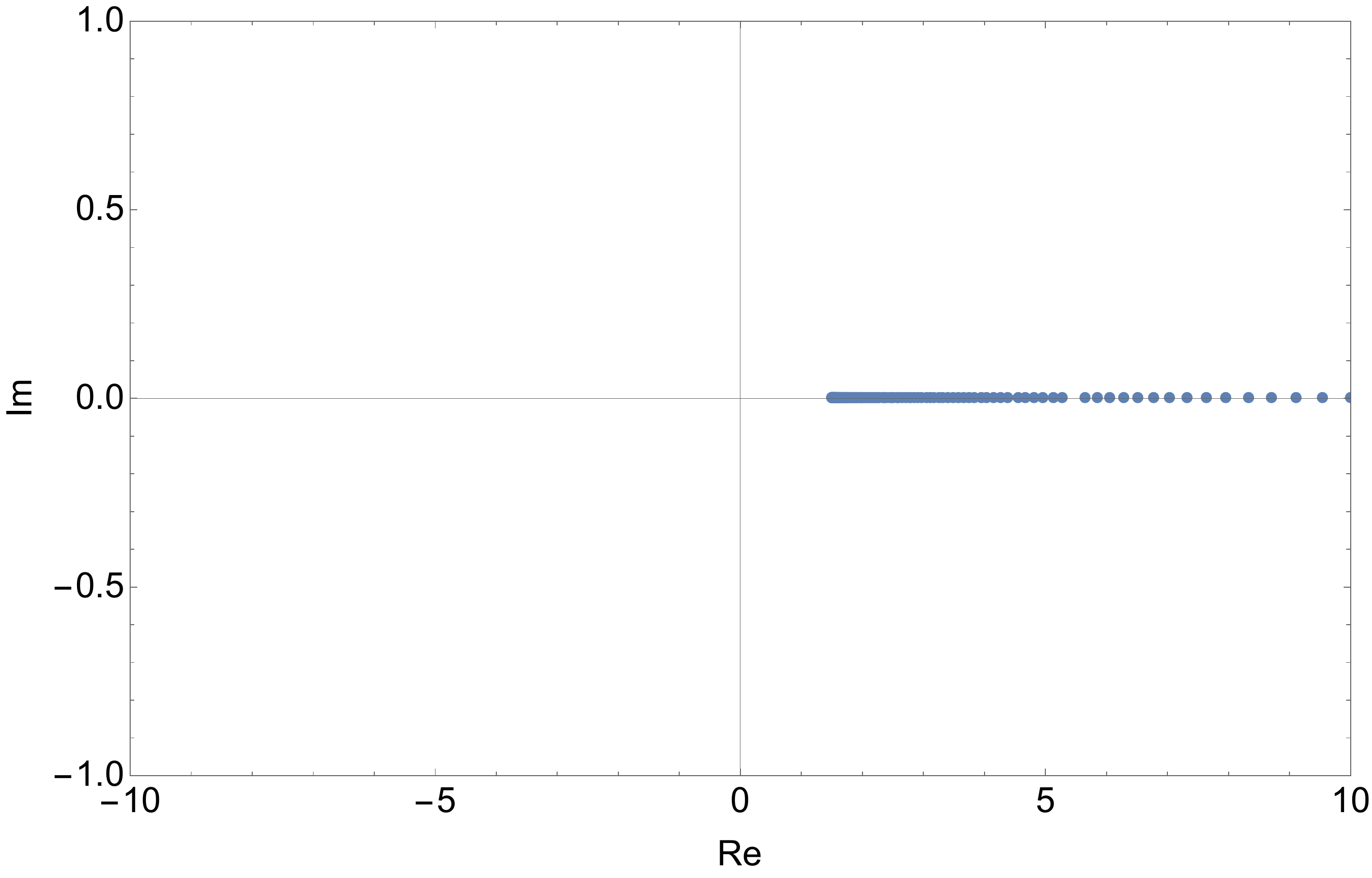}
    \caption{Singularity structure of the Borel transform \Eq{eq:padesum} from Pad\'e approximation. All poles are on the real axis, implying a brunch cut from the one closest to the origin in located at $3/2$ to $+\infty$. A similar figure can be found in \cite{Basar:2015ava}.}
    \label{fig:poleBRSSS}
\end{figure}

One way to reveal the hidden information in the gradient expansion, especially the emergence of non-hydrodynamic contributions from the hydrodynamic equation of motion, is to apply Borel resummation technique. With respect to the hydrodynamic gradient expansion \Eq{eq:BRSSS_ge}, Borel transform defines a new series as
\be
\label{eq:padesum}
{\cal B}[g_{\rm hydro} (z)]\equiv 
\sum_{n=0}^\infty \frac{f_n}{n!} z^n \,.
\ee
With an extra factor of $1/n!$ introduced, this series now has a finite radius of convergence. It can be shown that this new convergent series is related to the original hydrodynamic gradient expansion via an inverse Laplace transform, so that the Borel resummation of the hydrodynamic gradient expansion is obtained as 
\begin{align}
\label{eq:Boreltransf}
\tilde g_{\rm hydro}(w) =& \sum_{n=0}^\infty  \frac{f_n}{ w^n}\times
\frac{1}{n!}
\int_0^\infty dz e^{-z} z^{n}\cr
=&\int_0^\infty dz e^{-z} \sum_{n=0}^\infty \frac{f_n}{n!} \left(\frac{z}{w}\right)^n \cr
=& w \int_0^\infty dz e^{-zw} {\cal B}[g_{\rm hydro}(z)]\,.
\end{align}
One may check that $\tilde g_{\rm hydro}(w)$ is a solution to the nonlinear differential equation \Eq{eq:gBRSSS}. Without singularities, the integral can be carried out in a straightforward manner. 
For an asymptotic series, such as the hydrodynamic gradient expansion we are considering, however, singularities in the Borel transform are expected on the real axis, which give rise to additional contributions.

In \Fig{fig:poleBRSSS} the singularity structure of the Borel transform is shown on the complex Borel plane, where a series of poles on the real axis are observed. This structure is estimated numerically by a symmetric Pad\'e approximation of the Borel transform up to truncation order $n=300$. Note that, the leading pole (the pole closest to the origin) lies at $z=S$, is not quite sensitive to truncation orders. Ideally, the Borel transform of the original asymptotic series should result in a brunch cut on the real axis starting from $z_0$, as indicated by the accumulated poles in \Fig{fig:poleBRSSS}. To avoid the brunch cut on the real axis, the integral contour connecting zero and infinity in \Eq{eq:Boreltransf} needs to be analytically continued to the complex plane. Upon the integration contour considered above or below the real axis, there exists a complex ambiguity. This ambiguity results in the hydrodynamic gradient expansion a complex term. With respect to the leading pole, ${\cal B}[g_{\rm hydro}(z)]\sim(z-S)^{-\beta}$, with $S$ and $\beta$ real constants given in \Eq{eq:sbeta}, the complex ambiguity leads to
\be
{\rm Im} [\t g_{\rm hydro} (w)]\sim \pm \pi e^{-S/w} w^{\beta}\,.
\ee
Because the solution $g(w)$ is real and definite, this complex ambiguity in $g_{\rm hydro}$ must be cancelled by a term with the same exponentially suppressed factor. That is to say, the singularity of Borel resummation and the reality condition implies the existence of an extra contribution to the hydro gradient expansion, which is exponentially suppressed.

The existence of such extra terms with exponential factors can be proved as well through a small perturbation around the hydro gradient expansion. In \Eq{eq:gBRSSS}, assuming $g(w) = g_{\rm hydro}(w) + \delta g(w)$, one indeed finds an equation for $\delta g(w)$, to be solved asymptotically by, 
\be
\delta g(w)\sim e^{-Sw } w^{\beta} 
\ee

In fact, the complex ambiguity arises not only from the leading singularity, and cancellation of all the complex ambiguities requires the complete form of solution be a trans-series, rather than a simple power series expansion,
\be
g(w) = \sum_{n=0} (\sigma e^{-Sw} w^\beta)^n g^{(n)}(w)\,,
\ee
where obviously, the leading order gives $g^{(0)}(w)=g_{\rm hydro}(w)$.
In the trans-series, $\sigma$ is complex parameter denoting the order of trans-series expansion, whose real part is related to initial condition. At each order, 
the function $g^{(n)}(w)=\sum_k f^{(n)}_k /w^k$ itself is an asymptotic series, which with respect to the Borel resummation gives rise to a complex ambiguity to be cancelled by the similar factor from the next order term in the trans-series. The imaginary part of $\sigma$ is determined as a consequence of the cancellations. This is the typical property of the resurgent theory~\cite{Aniceto:2018bis,Aniceto:2013fka}, that different orders in the trans-series are related via the cancellation of complex ambiguities. Mathematically, it is not a surprise to expect the trans-series solution with exponentially suppressed factor for a nonlinear first order differential equation.

For asymptotically large $w$, the higher order transient contributions in the trans-series are suppressed. However, in the small $w$ regime, i.e., the non-perturbative regime of hydrodynamic gradient expansion, these higher order contributions become important. This can be shown, for instance, through the reconstruction of the attractor solution from the Borel resum of the trans-series solution. A key step in the 
procedure is the identification of the real part of the $\sigma$ parameter, corresponding to the initial condition that determines the attractor solution, which as we discussed before, corresponds to $g_{\rm hydro}(0^+) \to g_+$. Given ${\rm Re} \sigma$, one needs to resum the trans-series order by order, according to the detailed cancellation rules provided by the resurgence relations. For a conformal fluid it has been shown numerically that attractor does emerge, provided higher orders in the trans-series are taken into account~\cite{Heller:2015dha}. For some of the fluid systems with analytical solutions, the procedure can be proved explicitly by Borel resum of the trans-series to infinite orders~\cite{Blaizot:2020gql}.

\subsubsection{From kinetic theory}

The discussion in the previous section relies on an equation of motion provided from hydrodynamics, where viscous corrections are introduced up to the second order. Although a series expansion can be generated from the equation to arbitrary orders, based on the equation with second order viscous (gradient) corrections, this series expansion does not capture consistently the entire information of the off-equilibrium physics. For instance, it would not be surprising to realize that in \Eq{eq:BRSSS_ge} the expansion coefficients with $n\ge3$, are modified once in the original hydrodynamic equation of motion, third order, and higher order viscous corrections are taken into account~\cite{Blaizot:2019scw}. To formulate a gradient series that is compatible with the off-equilibrium system evolution, one has to solve the full transport equation~\cite{Heller:2016rtz,Heller:2018qvh,Strickland:2018ayk,Blaizot:2017ucy}.

With respect to the Bjorken symmetry, the general form of the Boltzmann equation~\cite{groot1980relativistic},
\be
p^\mu \partial_\mu f + \Gamma^{\alpha}_{\mu\nu} p^\mu p^\nu \frac{\partial}{\partial p^\alpha} f
= \hat{\cal C}[f]
\ee
is simplified. Especially, in the Milne coordinates, in the slide of $z=0$ (or $\xi=0$), the left hand side of the kinetic equation reduces to
\be
p^\mu \partial_\mu f + \Gamma^{\alpha}_{\mu\nu} p^\mu p^\nu \frac{\partial}{\partial p^\alpha} f
\,\to\,
p^0\left(\frac{\partial }{\partial \tau}- \frac{p_z}{\tau} \right)f \,,
\ee
where the phase-space distribution becomes a function of $(\tau,\p)$. As expected, this corresponds to the same kinematic domain as in the fluid dynamics in the previous section. We further consider a relaxation time approximation for the collision kernel, so that the kinetic equation is further simplified,
\be
\label{eq:simpleKin}
\left(\frac{\partial }{\partial \tau}- \frac{p_z}{\tau} \right)f(\tau,\p)
= -\frac{f(\tau,\p)-f_{\rm eq}(\tau,\p)}{\tau_R}\,.
\ee
There exists a formal and analytical solution to \Eq{eq:simpleKin}. For a relaxation time with arbitrary $\tau$-dependence, $\tau_R(\tau)$, the formal solution is~\cite{BAYM198418,PhysRevC.88.024903},
\begin{widetext}
\be
f(\tau,\p) = D(\tau,\tau_0) f(\tau_0,\p_\perp,p_z \tau/\tau_0) 
+ \int_{\tau_0}^\tau \frac{d\tau'}{\tau_R(\tau')} D(\tau,\tau') f_{\rm eq}(\sqrt{
\p_\perp^2+(p_z \tau'/\tau_0)^2}/T(\tau'))\,.
\ee
\end{widetext}
In this solution, $f_{\rm eq}$ is the equilibrium distribution, as a function of temperature which is to be fixed via the Landau matching condition: $\epsilon=\epsilon_{\rm eq}\propto T^4$. Apparently, as a consequence of conformal symmetry, $f_{\rm eq}$ does not depend on chemical potential. The function appearing in the first term is the free-streaming solution (when collision kernel vanishes) of the kinetic equation $f_{\rm FS}(\tau,\p) = f(\tau_0, \p_\perp, p_z \tau/\tau_0)$, and the time evolution function
\be
D(\tau_2,\tau_1) \equiv \exp\left[-\int_{\tau_1}^{\tau_2} d\tau' \frac{1}{\tau_R(\tau')}\right]\,.
\ee

Conservation of energy-momentum is implied in the kinetic equation,
$\partial_\mu T^{\mu\nu} = 0$,
where the energy-momentum tensor is defined in kinetic theory as,
\be
T^{\mu\nu} = \int \frac{d^3 \p}{E_p} p^\mu p^\nu f_\p\,.
\ee
In the case of Bjorken symmetry, the independent components of the energy-momentum tensor are the diagonal components, which include the local energy density,
\be
\epsilon = T^{00} \equiv \int d^3 \p E_p f(\tau, \p)\,,
\ee
the longitudinal pressure 
\be
\P_L = T^{zz} \equiv \int \frac{d^3 \p}{E_p} p_z^2 f(\tau, \p)\,,
\ee
and the transverse pressure,
\be
\P_T = T^{xx} = T^{yy} \equiv \frac{1}{2}
\int \frac{d^3 \p}{E_p} p_\perp^2 f(\tau, \p)\,.
\ee
Note that with respect to the conformal symmetry, $\epsilon=\P_L+2\P_T$. In terms of these components, the conservation of energy and momentum gives,
\be
\label{eq:kinEcon}
\tau \frac{d\epsilon }{d\tau} + \frac{4}{3}\epsilon + \frac{2}{3}(\P_L-\P_T)=0\,.
\ee
From the above equation, one notices that ideal hydrodynamic equation of motion is recovered when pressures are isotropized, $\P_L=\P_T$. In the case of viscous hydrodynamics, the pressure anisotropy corresponds to small viscous corrections. With respect to the BRSSS form of viscous hydrodynamics, it is~\cite{Blaizot:2017lht}
\be
\P_L-\P_T = - \frac{2\eta}{\tau} + \frac{4}{3\tau^3} (\lambda_1-\eta \tau_\pi) + O\left(\frac{1}{\tau^3}\right)\,.
\ee

\begin{figure}
    \centering
    \hspace*{-10mm}\includegraphics[width=.95\linewidth]{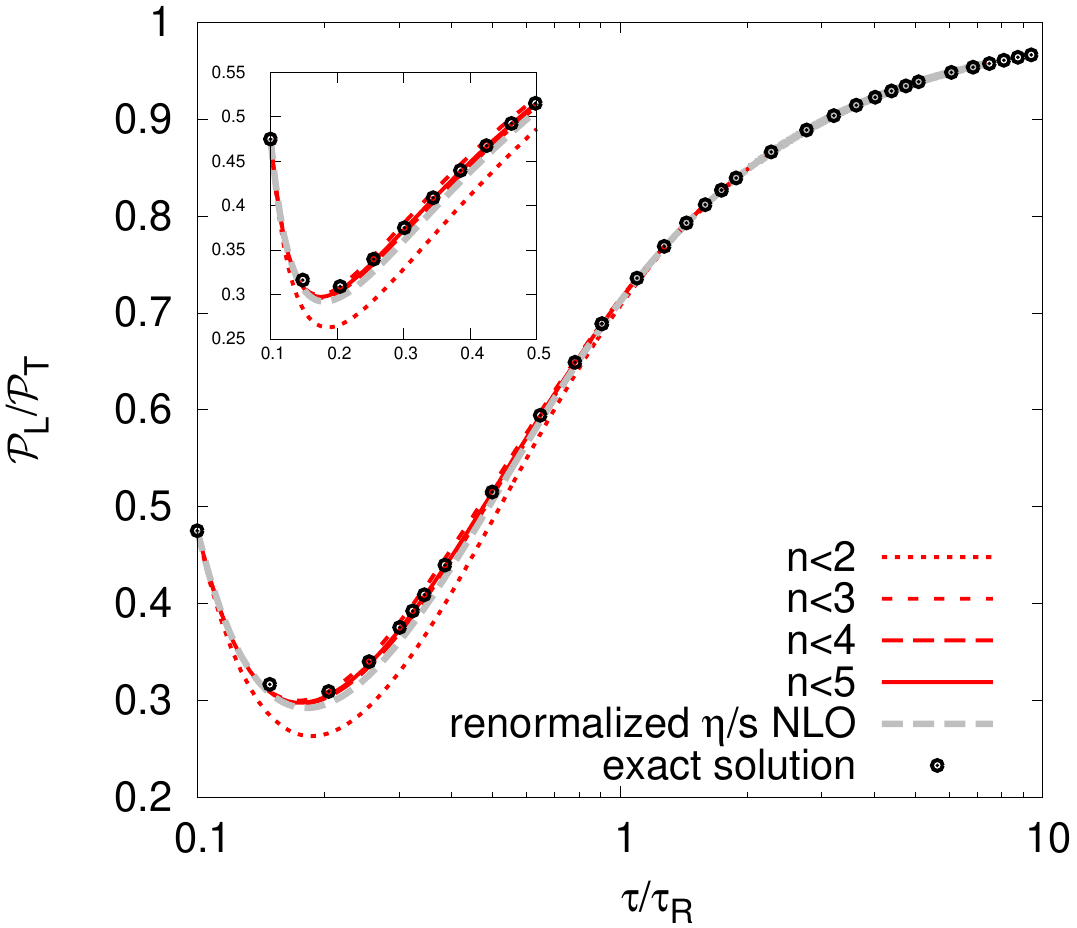}
    \caption{Pressure anisotropy solved from the truncated moment equations with a truncation including two, three, four and five $\L$-moments, in comparison with the exact solution of the kinetic equation (open symbols). Solution with respect to viscous hydrodynamics, or the truncated moment equation at $\L_0$ and $\L_1$, but using a renormalized $\eta/s$ is shown as the gray dashed line.
    \label{fig:truncatedL}}
\end{figure}

\subparagraph{The $\L$-moments.} The energy-momentum tensor $T^{\mu\nu}$ belongs to one specific set of moments of the phase-space distribution. We define the ${\cal L}$-moment,
\be
{\cal L}_n = \int d^3 \p E_p P_{2n} (p_z/E_p) f(\tau,\p)
\ee
using the Legendre polynomial $P_n(x)$. As a result of the parity symmetry in the Bjorken expansion, moments associated with odd orders of the Legendre polynomials vanish. In addition to the Legendre polynomial that specifies asymmetry in the phase space, the weight $E_p$ is chosen such that the ${\cal L}$-moments are of the same dimension as the energy-moment tensor. Indeed, it is straightforward to verify that
\begin{align}
&\L_0=T^{00}=\epsilon=\P_L+2\P_T\,,\nonumber\\
&\L_1=T^{\perp\perp}-T^{zz} = \P_L-\P_T\,,
\end{align}
hence the pressure anisotropy can be expressed in terms of the $\L$-moments as
\be
\label{PLoverPT}     
\frac{\P_L}{\P_T}=\frac{\L_0+2\L_1}{\L_0-\L_1}\,.
\ee
Legendre polynomials provide a complete set of decomposition in the angular dependence with respect to Bjorken symmetry, but the reconstruction of the phase-space distribution requires also a complete mode decomposition for the $E_p$ dependence. For instance, the $E_p$-dependence in $f(\tau,\p)$ can be decomposed by Larguerre polynomials~\cite{Bazow:2016oky}. 
Although generalized moments of the distribution function can be introduced (cf. Ref~\cite{Denicol:2012cn,Tinti:2018qfb,Strickland:2018ayk}), the $\L$-moments are sufficient for the description of system hydrodynamization, concerning especially the evolution of pressures and energy density. For instance, with respect to conformal symmetry, $\L_0$ and $\L_1$ fully determine the components in the energy-momentum tensor $T^{\mu\nu}$. With higher order $\L$-moments included, the description of system evolution can be even improved. An illustration will be given later in \Fig{fig:truncatedL}. 

It is also interesting to notice that, the coupled equations for $\L_0$ and $\L_1$ are analogous to the so-called anisotropic hydrodynamics (ahydro)~\cite{Martinez:2010sc,Florkowski:2010cf}, where the pressure difference is considered as an individual field for out-of-equilibrium fluids. In a similar manner, higher order $\L$-moments play the role of viscous corrections, in comparison to the viscous anisotropic hydrodynamics (vahydro)~\cite{Bazow:2013ifa}. 

With respect to the analytical formal solution of the distribution function, the analytical solution of $\L$-moments can be obtained,
\begin{widetext}
\be
\label{eq:formalL}
\L_n(\tau) = D(\tau,\tau_0) \L_n^{(0)}(\tau) 
+ \int_{\tau_0}^\tau \frac{d\tau'}{\tau_R(\tau')} D(\tau,\tau') \L_0(\tau')
(\tau'/\tau) \F_n(\tau')
\ee
\end{widetext}
where the function $\F_n(x)$ is defined as
\begin{align}
\label{calFdef}
\mathcal{F}_{n}(x)\equiv\frac{1}{2}\int_{-1}^1 dy &\left[1-(1-x^2)y^2\right]^{1/2}\cr
\times &P_{2n}\left(\frac{xy}{\left[1-(1-x^2)y^2\right]^{1/2}}\right)\,.
\end{align}
Note that in the limit of $x\to0$, $\F_n(0)\to \pi P_{2n}(0)/2$, and in the limit $x\to 1$, $\L_{n\ne0}(1)=0$. For the case of $n=0$, the above integral can be analytically evaluated, resulting in
\be
\F_0(x)=\frac{1}{2} \left(x-\frac{i \cosh ^{-1}(x)}{\sqrt{1-x^2}}\right)\,.
\ee
The first term in \Eq{eq:formalL} contains the free-streaming moment $\L_n^{(0)}(\tau)$, which is given by
\be\label{FSLnISO}
\L_n^{(0)}(\tau)=\epsilon_0 \,\frac{\tau_0}{\tau}\, \F_n\left( \frac{\tau_0}{\tau}\right)\,,
\ee 
with $\epsilon_0$ being the initial energy density. \Eq{eq:formalL} should be regarded as an integral equation, for which $\L_0$ must be solved with respect to the initial condition $\epsilon_0$. Once $\L_0(\tau)$ is provided, higher order $\L_n$ can be determined accordingly. 

\Eq{eq:formalL} allows one to study the solution of energy density in power of $1/w=\tau_R/\tau$, \ie, the gradient expansion at late time. For instance, the gradient expansion can be generated using integration by parts in the integral equation of $\L_0$, which again leads to an asymptotic series~\cite{Heller:2018qvh}. With the asymptotic series given, in analogous to the hydrodynamic gradient expansion, the emergence of a trans-series solution, and hence resurgence phenomenon, etc., can be obtained following the standard procedure of Borel resum. 

Alternatively, one may start from a set of coupled equations of the $\L$-moments. By substituting the definition of $\L$-moments to the kinetic equation \Eq{eq:simpleKin}, one finds a first order differential equations where adjacent $\L$-moments are coupled,
\be
\label{eq:eomL}
\partial_\tau \L_n + \frac{a_n\L_n+b_n \L_{n-1}+c_n \L_{n+1}}{\tau}
=-\frac{\L_n}{\tau_R}(1-\delta_{n0})\,,\cr
\ee
where $n=0,1,2,\ldots$.
The constant coefficients $a_n$, $b_n$ and $c_n$ arise from the recursion relation of Legendre polynomials,
\beq
a_n &=&
\frac{2(14 n^2+7n-2)}{(4n-1)(4n+3)}\,,\quad
b_n=\frac{(2n-1)2n(2n+2)}{(4n-1)(4n+1)}\,,\nonumber\\
c_n&=&\frac{(1-2n)(2n+1)(2n+2)}{(4n+1)(4n+3)}\,,
\eeq
reflecting the geometric nature of Bjorken expansion. These can be also understood as the limiting case of the Clebsch–Gordan coefficients without couplings between transverse and parity-odd modes. The first several of these constants are
\beq
&&a_0=4/3,\quad b_1=0,\quad c_0=2/3,\nonumber\\
&&a_1=38/21,\quad b_2= 8/15, \quad c_1=-12/35. 
\eeq
The equation for $\L_0$ is exactly the conservation of energy-moment, as in \Eq{eq:kinEcon}, while higher order $\L_n$'s bring in corrections. In the vicinity of local equilibrium, these are the viscous corrections.

To solve the time evolution of these moments, in comparison with the exact solution from the kinetic equation, one would expect truncating the coupled equations at a finite order as a good approximation. In \Fig{fig:truncatedL}, the numerically solved pressure anisotropy $\P_L/\P_T$ is plotted as a function of $\tau/\tau_R$, for the case of a conformal medium that $\tau_R T=$ constant with respect to one specified initial condition, $\P_L/\P_T\approx 0.49$. The exact solution to the kinetic equation is shown as open symbols, comparing to which, the simplest truncation of the moment equations at order $n<2$ involving moments $\L_0$ and $\L_1$ (dotted line), already captures the bulk property of time evolution. Deviations from the lowest order approximation are significant only when $\tau/\tau_R<1$, but negligible in the hydrodynamic regime when $\tau/\tau_R\gg1$. With higher order $\L$-moments involved, improvement from these corrections is appreciated. With the truncation at $n<5$ (with $\L_0,\ldots,\L_4$ being taken into consideration), a satisfactory description of the pressure anisotropy in the whole time evolution is obtained.

In fact, effectiveness of the truncation of the moment equations are guaranteed owing to the existence of fixed points, and can be as studied analytically in the limiting cases. If one considers the free-streaming, or to say, focuses on the very early time limit $\tau/\tau_R\to0$ that collisions are effectively suppressed by expansion, the solutions of moments are analytical. This is even obvious in the formal solution, \Eq{eq:formalL}. One may also recast the set of moment equations into a matrix form, with respect to the vector,
\be
\psi = (\L_0,\L_1,\L_2,\ldots)\,.
\ee
Correspondingly, the dynamics of free-streaming is captured by a constant tri-diagonal matrix, 
\be
\hat H = 
\begin{pmatrix}
  a_0 & c_0 & 0 & 0   & \ldots\\
  b_1 & a_1 & c_1 & 0 & \ldots\\
  0 & b_2 & a_2 & c_2 & \ldots\\
  \ldots&\ldots&\ldots&\ldots&\ldots\\
 \end{pmatrix}\,,
\ee
so that the equation of free streaming becomes,
\be
\partial_\rho \psi + \hat H \psi = 0\,.
\ee
We have  introduced for convenience $\rho = \ln (\tau/\tau_0)$. The solution of moments can be found provided the eigenvalues of the matrix are determined. We note that eigenvalues of $\hat H$ are complex, with eigen-vectors satisfying,
\be
\hat H \phi_n = \lambda_n \phi_n\quad \to \quad
\phi_n(\tau) = \phi_n(\tau_0) e^{-\lambda_n \rho}\,.
\ee
Let us order these eigenvalues by their real part, \ie,
\be
{\rm Re}\lambda_0 < {\rm Re}\lambda_1 < {\rm Re}\lambda_2 <  \ldots < {\rm Re}\lambda_\infty\,,
\ee
then the solution of moments is,
\begin{align}
\psi(\tau) &= \sum_n \kappa_n e^{-\lambda_n \rho} \phi_n(\tau_0)\nonumber\\
&=e^{-\lambda_0 \rho} \left[\kappa_0 \phi_0(\tau_0) + \sum_{n\ne0} \kappa_n e^{-(\lambda_n - \lambda_0)\rho}\phi_n(\tau_0)\right]\nonumber\\
&\xrightarrow[]{} e^{-\lambda_0\rho} \kappa_0 \phi_0\,,
\end{align}
where $\kappa_n$ are constant coefficients fixed by the initial condition. Eventually, the evolution of moment is dominated by the ground-state mode, on a time scale $(\lambda_n-\lambda_0)\rho\sim\Delta\lambda \rho \gg 1$, irrespective of initial conditions~\cite{Heller:2018qvh}. The gap of eigen-values is order unity, $\Delta\lambda\sim 1$. 

\begin{figure}
    \centering
    \hspace*{-10mm}\includegraphics[width=.95\linewidth]{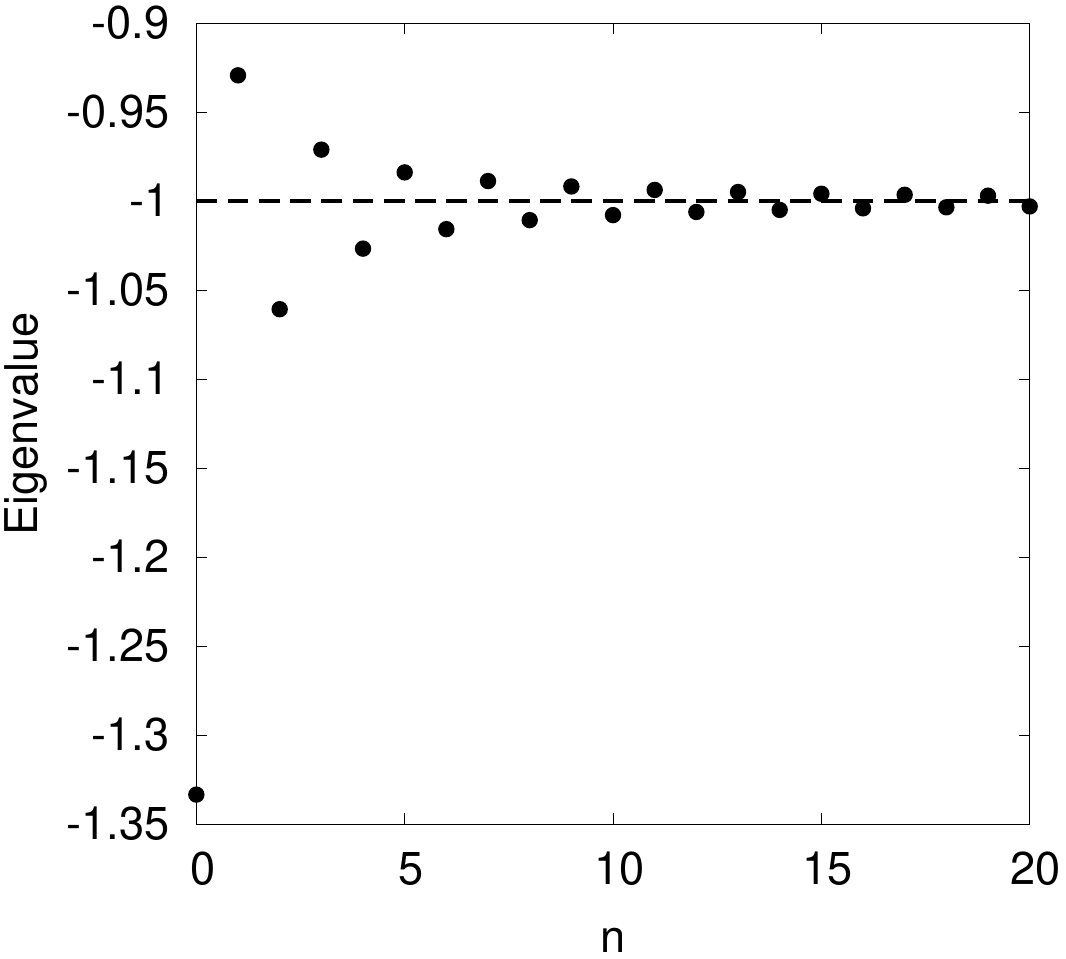}
    \caption{Ground-state eigen-value of the matrix $\hat H$, with respect to truncation orders. The asymptotic value is 1, while $\lambda_0\approx-0.929$ when truncating at $n<2$. Figure taken from \cite{Blaizot:2019scw}.
    \label{fig:lambda0}}
\end{figure}

If one generalizes the definition of the function $g(w)$ in \Eq{eq:g0def}, for all the moments,
\be
\label{eq:gdef}
g_n(w) \equiv \frac{\rmd \ln\L_n}{\rmd \ln\tau}\,,
\ee
the dominance of the ground-state indicates the existence of a stable fixed point, $g_n(w) \to -\lambda_0$, regardless of the order $n$. In the similar way, an unstable fixed point is also expected, corresponding to $\lambda_\infty$. This is very similar to the observations from the fluid dynamics analysis, albeit that both the stable free-streaming fixed point and the unstable fixed point depends weakly on the truncation order. For truncation at $n<2$, stable fixed point is $g_+\approx -0.929$, while the unstable fixed point is $g_-\approx-2.213$, in analogous to fixed point analysis from fluid dynamics. This is not accidental, but a direct consequence that the simplest truncation of moment equations leads to second order viscous hydrodynamics. In \Fig{fig:lambda0}, the ground-state eigenvalue is plotted as a function of truncation order. Although asymptotically when $n\to\infty$, $\lambda_0\to -1$, truncating at finite orders only gives a small correction. This observation guarantees the validity of moments truncation in the free-streaming limit. Corresponding to the stable fixed point, or the ground state eigen-value, the ground-state eigen-vector is determined that fixes the ratios between moments,
\be
\phi_0:\quad \L_n = P_{2n}(0) \L_0 = (-1)^n \frac{(2n-1)!!}{(2n)!!} \L_0\,.
\ee
In the original phase-space distribution, these $\L$-moments with the specified ratios characterizes a distribution spanning along $p_\perp$ and shrinking in $p_z$. One may also prove that when truncation order goes to infinity, the unstable fixed point is $g_-\to -2$, by noticing that $a_n+b_n+c_n = 2$.
The corresponding eigen-vector is given by 
\be
\phi_{\infty}:\quad
\L_0=\L_1=\L_1=\ldots
\ee

In the opposite limit with $\tau/\tau_R\to\infty$, the moments admit a series expansion in powers of $1/\tau$,
\be
\L_n = \sum_{m=n}^\infty \frac{\alpha_n^{(m)}}{\tau^n}\,.
\ee
This structure follows from the Chapman-Enskog expansion of the kinetic theory~\cite{Blaizot:2017lht}. 
Except for $\L_0$, $\alpha_n^{(0)}=\epsilon \delta_{n0}$, the expansion coefficients are directly related to those transport coefficients in viscous hydrodynamics. For instance, by noticing that the leading term in $\L_1$ satisfies $\L_1=\P_L-\P_T\propto\eta/\tau$, one is able to identify the shear viscosity as
\be
\label{eq:etaL}
\eta = \frac{b_1}{2} \tau_R \epsilon\,.
\ee
Coefficients in the expansion are dimensionful, which can be further expressed in terms of dimensionful variables in the original moment equations, $\alpha_n^{(m)}=B_n^{(m)} \epsilon \tau_R^n$, with $B_n^{(m)}$ dimensionless. The leading order and the next-leading order expansion coefficients can be solved analytically. The leading order ones, $\alpha_n^{(n)}$, in particular, determine the asymptotic value of $g_n(w)$. For instance, for a constant relaxation time, taking into account the time dependence of energy density in ideal fluid, $\epsilon\sim \tau^{-4/3}$, one finds $\L_n\sim \tau^{-4/3-n}$. For a conformal system where $\tau T=$const., on the other hand, one finds $\L_n\sim \tau^{-4/3-2n/3}$. Therefore,
\be
\label{eq:gnasymp}
g_n(\infty)=\begin{cases}
-\frac{4}{3} - n\,,\qquad \tau_R = \mbox{const.}\\
-\frac{4}{3} - \frac{2n}{3}\,,\quad \tau_R T = \mbox{const.}
\end{cases}
\ee

\begin{figure}
    \centering
    \hspace*{-10mm}\includegraphics[width=.95\linewidth]{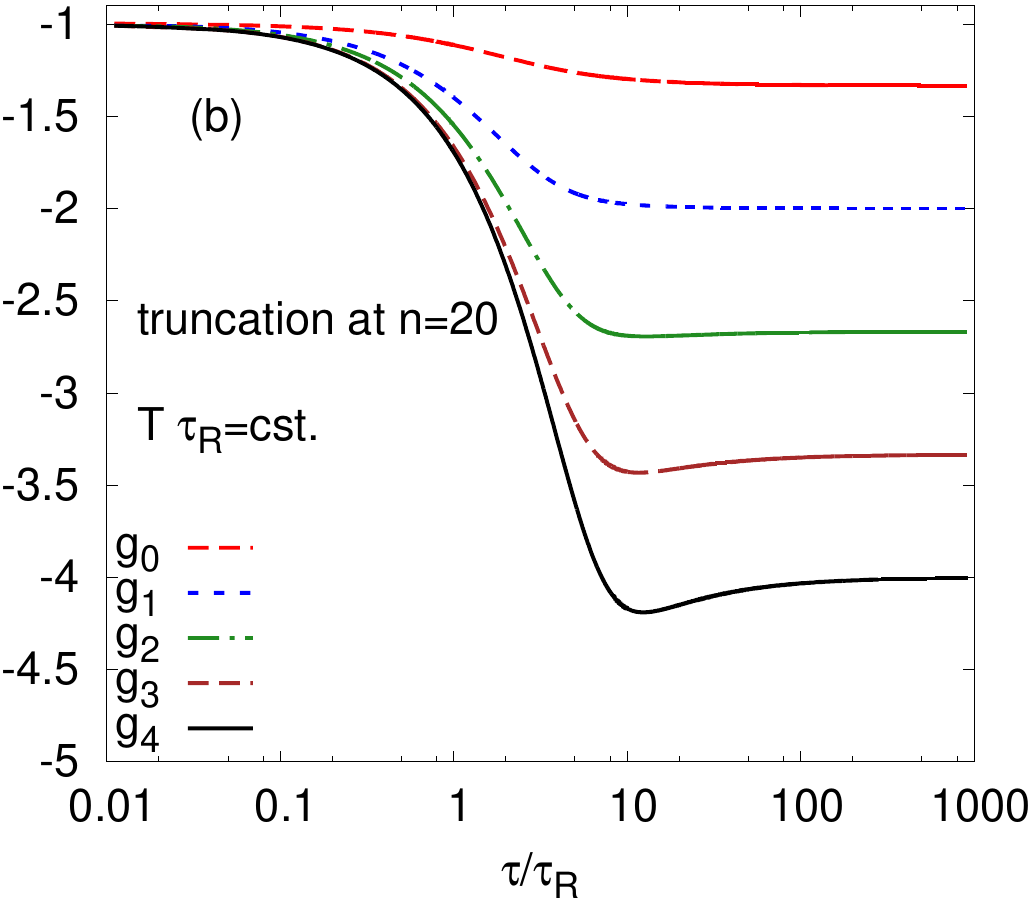}
    \caption{Attractor solution of $g_n$ for $n=0,1,\ldots,4$, calculated from the numerical solution of the coupled moment equations with a truncation order at $n=20$. A conformal system with $\tau_R T=$const. is considered, for which the asymptotic values of $g_n(w)$ in the hydrodynamic regime depends on order $n$, according to \Eq{eq:gnasymp}. Figure taken from Ref.~\cite{Blaizot:2019scw}.
    \label{fig:attractorLn}}
\end{figure}

These asymptotics represent the hydrodynamic fixed points of the moments of different orders, which the moments would eventually approach, irrespective of initial conditions. Correspondingly, attractors are smooth solutions that connect from the free streaming fixed point and the hydrodynamic fixed point. Since the hydrodynamic fixed points differ in orders for different $\L_n$, there are consequently infinite number of attractors from the coupled moment equations, or the original kinetic equations. This is also noticed in other forms of moments as well~\cite{Strickland:2018ayk}. In \Fig{fig:attractorLn}, the attractors of the first five order of $\L$-moments are plotted in terms of $g_n(w)$, for a conformal system. 

\subparagraph{The $\L$-moments and variants of hydrodynamics.} 
As we have presented, the conservation of energy and momentum is only a subset of the coupled moment equations, actually the leading one. Noting that conservation of energy and momentum $\partial_\mu T^{\mu\nu}=0$ involves the first two moments $\L_0$, $\L_1$. The simplest truncation that solves the conservation of energy and momentum is
\begin{subequations}
\label{eq:simpletrun}
\begin{align}
\label{eq:sta}
&\partial_\tau \L_0 + \frac{1}{\tau}(a_0 \L_0 + c_0 \L_1) = 0 \\
\label{eq:stb}
&\partial_\tau \L_1 + \frac{1}{\tau}(a_1 \L_1 + b_1 \L_1) = - \frac{1}{\tau_R} \L_1\,,
\end{align}
\end{subequations}
Together with the traceless condition $T^{\mu}_\mu=0$, all components in $T^{\mu\nu}$ can be thus determined. 

In the hydrodynamic regime, with $\tau/\tau_R\to\infty$, one would expect \Eqs{eq:simpletrun} to be identified as the hydrodynamic equations of motion. In fact, \Eq{eq:sta} is just the conservation of energy and momentum, $\partial_\mu T^{\mu\nu}=0$, for a Bjorken expanding system. \Eq{eq:stb}, on other hand, generalizes the constitutive relation that interprets the pressure anisotropy $\L_1=\P_L-\P_T$ in terms of viscous corrections. For the system with Bjorken expansion, at late time if one further identifies $c_0\L_1$ as the $\xi\xi$ component of the shear stress tensor, $c_0 \L_1 = \pi = \pi^\xi_\xi$, the truncated moment equations indeed leads to the Israel-Stewart hydro equations of motion,
\begin{subequations}
\begin{align}
\label{eq:mishydroa}
&\partial \epsilon + \frac{4}{3}\frac{\epsilon}{\tau} = -\frac{\pi}{\tau}\\
\label{eq:mishydrob}
&\pi= -\frac{4}{3}\frac{\eta}{\tau} - \tau_\pi \partial_\tau \pi - \beta_{\pi\pi}\tau_\pi \frac{\pi}{\tau}\,.
\end{align}
\end{subequations}
In obtaining the above equation, we have considered conformal EoS $\epsilon=3\P$ and used the shear viscosity from \Eq{eq:etaL}. By doing so, one finds the second order transport coefficient $\tau_\pi=\tau_R=5\eta/sT$. The transport coefficient $\beta_{\pi\pi}$ is precisely $a_1$, which agrees with~\cite{Denicol:2014ywa}. Because in Bjorken flow, ambiguity arises from interpreting $1/\tau$ as the $\xi\xi$ component of the shear tensor $\sigma_\xi^\xi$, or the expansion rate $\nabla\cdot U$, as indicated in \Eq{eq:bhydro}, the hydrodynamic constitutive relation from \Eq{eq:stb} is not unique. For instance, if one splits the last term in \Eq{eq:mishydrob} as
\be
\beta_{\pi\pi}\tau_\pi \frac{\pi}{\tau} = 
\tau_\pi\frac{4}{3} \frac{\pi}{\tau}
-\frac{3}{4}\left(\beta_{\pi\pi}-\frac{4}{3}\right) 
\frac{\tau_\pi}{\eta}\pi^2 + O\left(\frac{1}{\tau}\right)^3\,,\cr
\ee
the constitutive relation gives rise to the BRSSS formulation~\cite{Baier:2007ix},
\be
\pi = -\frac{4}{3}\frac{\eta}{\tau} - \tau_\pi\left(\partial_\tau \pi + \frac{4}{3}\frac{\pi}{\tau}\right) + \frac{3}{4}\left(\beta_{\pi\pi}-\frac{4}{3}\right)\frac{\tau_\pi}{\tau}\pi^2\,.\cr
\ee
Correspondingly, the second order transport coefficient $\lambda_1$ and $\beta_{\pi\pi}$ are related, and the evaluation in a weakly coupled medium
\be
\label{eq:betalambda}
\lambda_1 = \frac{3}{4}\left(\beta_{\pi\pi}-\frac{4}{3}\right) \eta\tau_\pi= \frac{5}{7}\eta\tau_\pi\,,
\ee
is confirmed~\cite{Dusling:2009df}.

\begin{figure}
    \centering
    \hspace*{-10mm}\includegraphics[width=.95\linewidth]{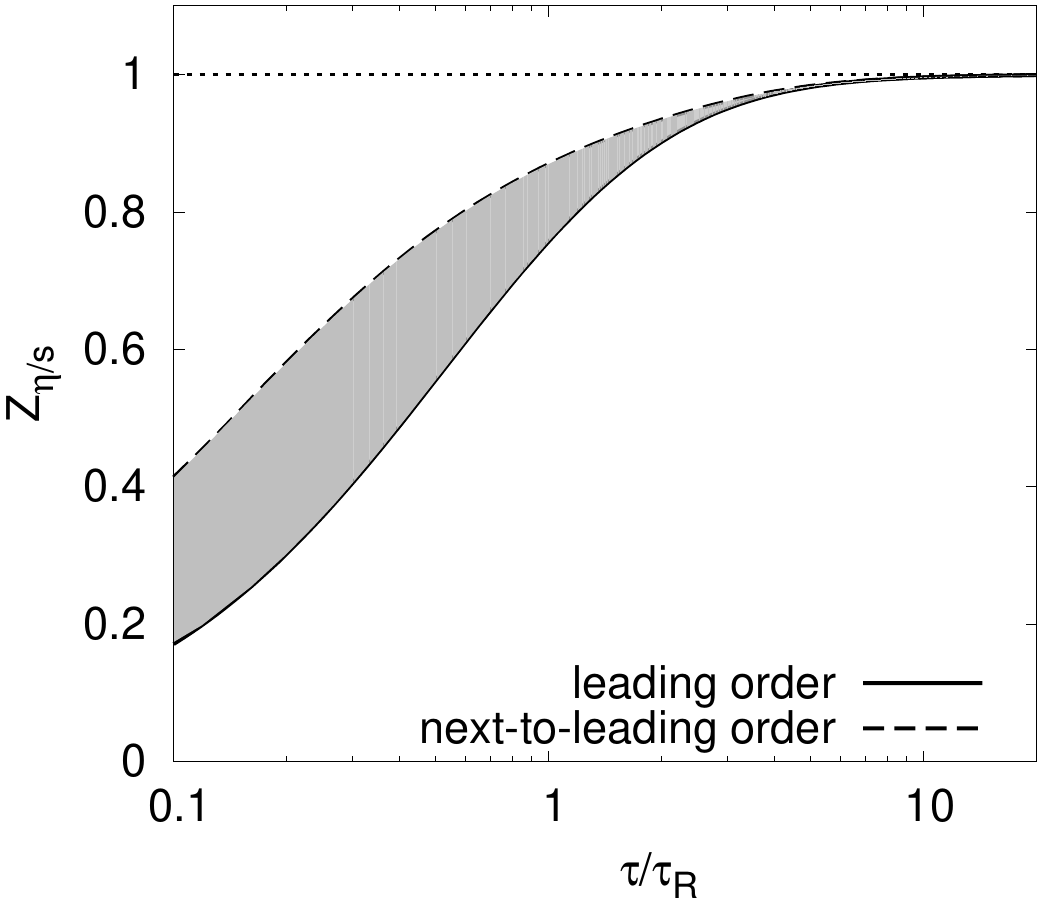}
    \caption{Renormalization of $\eta/s$ by out-of-equilibrium effects. Figure taken from Ref.~\cite{Blaizot:2017ucy}.
    \label{fig:Zetas}}
\end{figure}

Inclusion of higher order moments in the coupled equation will improve the quantitative characterization of the system evolution, as already been noticed. These corrections due to higher order moments can be explicitly incorporated in the equations even for the simplest truncation, by one additional term $\propto c_1\L_2/\tau$ in \Eq{eq:stb}~\cite{Blaizot:2019scw}. This term can be also absorbed into the collision term, via a redefinition of the relaxation time,
\be
\label{eq:renormZ}
\frac{1}{\tau_R}\to 
\left[1+\frac{c_1 \tau_R}{\tau}\frac{\L_2}{\L_1}\right]\frac{1}{\tau_R} 
\equiv  \frac{Z_{\eta/s}^{-1}}{\tau_R}\,.
\ee
It is then straightforward to see that, since the ratio in the factor is related to $g_2(w)$, 
\be
\frac{\L_2}{\L_1} = -\frac{b_2}{w + g_2(w) + a_2}\,,
\ee
one is allowed to solve the coupled moment equations to arbitrary orders, provided $g_2(w)$ is given precisely. In terms of the simplest truncation that involves dynamically only $\L_0$ and $\L_1$, as in the case of viscous hydrodynamics, \Eq{eq:renormZ} gives effectively a renormalized relaxation time $\tau_R$. Because $\eta/s\propto \tau_R$, this procedure equivalently renormalizes the ratio of shear viscosity to the entropy density, $\eta/s$, by a multiplicative renormalization factor $Z_{\eta/s}$. If one further considers the attractor as a generic generalization of hydrodynamic mode to out-of-equilibrium, and hence substitutes the attractor solution of $g_2(w)$ in calculations, the renormalized $\tau_R$ (or $\eta/s$) in an out-of-equilibrium system is obtained. In \Fig{fig:Zetas}, the factor $Z_{\eta/s}$ is obtained via the attractor solution of $g_2(w)$. Unless the system is close to equilibrium, $\tau/\tau_R\gg 1$, the out-of-equilibrium effects will reduce the value of $\eta/s$, making the out-of-equilibrium system more close to an ideal fluid~\cite{Lublinsky:2007mm,Romatschke:2017vte,Behtash:2018moe,Kamata:2020mka}. A numerical test of the renormalization scheme can be found in \Fig{fig:truncatedL}, where given a renormalized $\eta/s$, even the solution of the two-moment equations achieves good agreement in comparison with the exact solution.

\section{Phenomenological development}\label{sec.III}

Relativistic hydrodynamics, incorporated with lattice QCD based equation of state (EoS), viscosity, and initial state fluctuations, has been a precision tool to understand the dynamics of the strongly-coupled QGP and the experimental flow observables (see reviews \cite{Heinz:2013th, Gale:2013da, Yan:2017ivm}). Fluid dynamics serves as a universal long-wavelength description of the system's macroscopic degrees of freedom from the QGP to hadronic gas phase. 
This strongly-coupled description naturally breaks down as the system becomes increasingly dilute within its hadronic phase at low temperatures. One must then transit to a microscopic transport description. The numerical realizations of hadronic transport models are UrQMD \cite{Bass:1998ca,Bleicher:1999xi}, JAM \cite{Nara:1999dz}, and SMASH \cite{Weil:2016zrk}.
Such a hydrodynamics + hadronic transport hybrid theoretical framework has successfully described and even predicted various types of flow correlation measurements with remarkable precision \cite{Song:2010mg, Song:2011hk, Shen:2011eg, Heinz:2011kt}. 

In this section, we will review the recent phenomenological developments on modeling the full 3D dynamics at intermediate collision energies, current state-of-the-art constraints on the QGP transport properties, understanding collective behavior in small systems, and interdisciplinary cross-talks with other fields of science.

\begin{figure}[t!]
    \centering
    \includegraphics[width=0.95\linewidth]{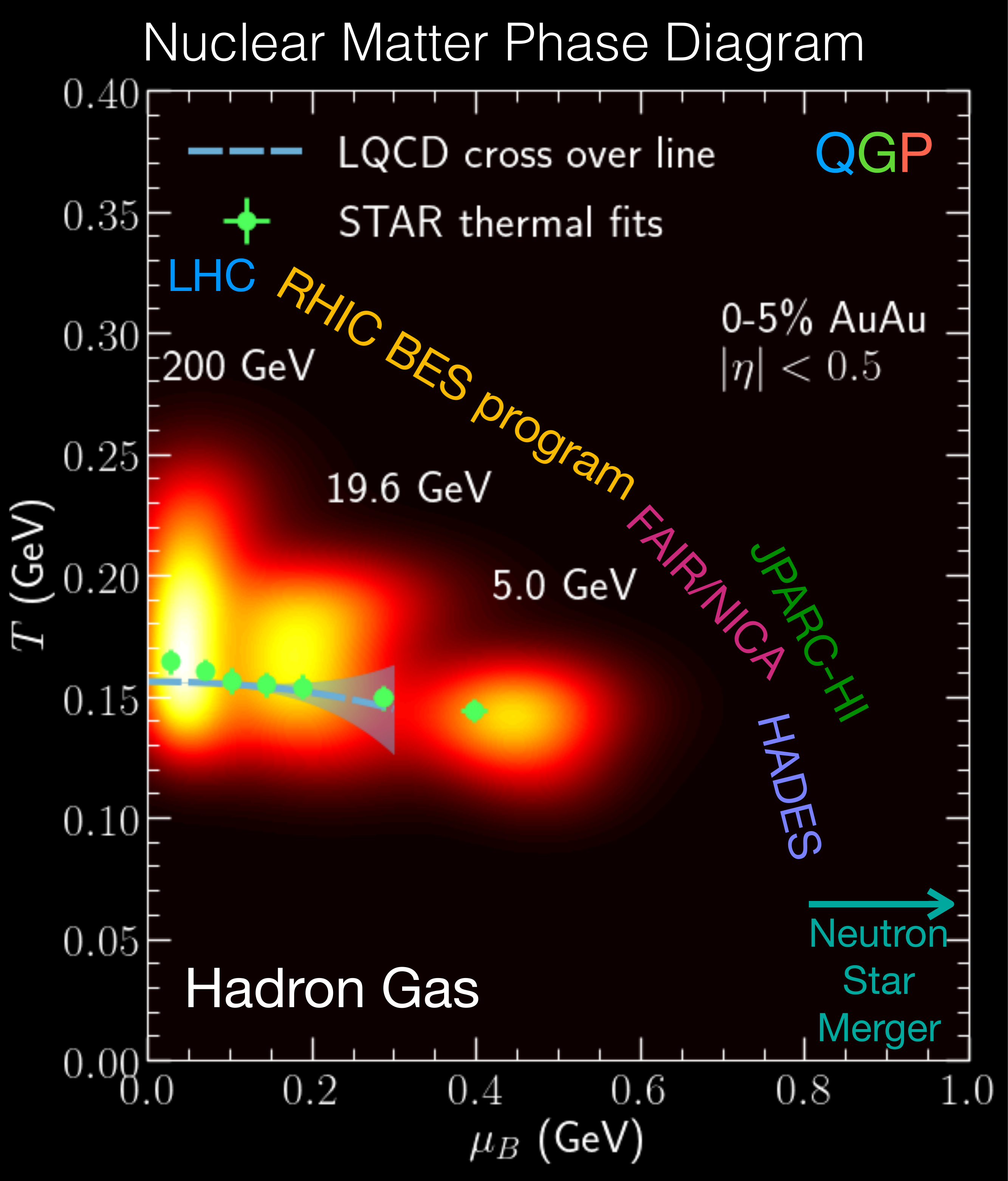}
    \caption{A sketch of the QCD phase diagram together with the current and future heavy-ion experimental programs. As relativistic heavy-ion collisions evolve from QGP to hadron gas phase, they explore the phase diagram of hot and dense QCD matter. Lattice QCD calculations identified a smooth cross-over between QGP and hadron gas for $\mu_B < 250$ MeV \cite{Bazavov:2018mes}. The chemical freeze-out points extracted from thermal fits at RHIC \cite{Andronic:2017pug,Adamczyk:2017iwn} are shown.
    The three blobs represent fireball trajectories of Au+Au collisions at RHIC BES energies mapped onto the QCD phase diagram event-by-event using the dynamical framework \cite{Shen:2017bsr}. Their brightness is proportional to the fireball space-time volume weighted by $T^4$. A cross-over phase transition is assumed in the simulations. This figure is adapted from \cite{Shen:2020gef}.}
    \label{fig:QCD_phase_diagram}
\end{figure}

\subsection{Hydrodynamic perspectives on Beam Energy Scan and longitudinal dynamics}\label{Sec:BES}

Quantifying the phase structure of QCD matter is one of the primary questions in relativistic heavy-ion physics.
First principles lattice QCD calculations have established that hadron resonance gas transits to the Quark-Gluon Plasma (QGP) phase as a smooth cross-over at vanishing net-baryon density \cite{Aoki:2006we}. In the meantime, many model calculations conjectured the presence of a first-order transition accompanied by a critical point at some finite net-baryon density in the QCD phase diagram (see e.g., \cite{Stephanov:2004wx, Bzdak:2019pkr} for a review).
Current heavy-ion experiment programs, such as the RHIC BES program \cite{Caines:2009yu, Mohanty:2011nm, Mitchell:2012mx, Odyniec:2015iaa}, the NA61/SHINE experiment at the Super Proton Synchrotron (SPS) \cite{Gazdzicki:2008kk, Abgrall:2014xwa}, as well as future experiments at the Facility for Antiproton and Ion Research (FAIR) \cite{Spiller:2006gj, Ablyazimov:2017guv}, Nuclotron-based Ion Collider fAcility (NICA) \cite{Sissakian:2009zza}, and JPARC-HI \cite{Sakaguchi:2019xjv}, routinely produce hot and dense QCD matter to probe an extensive temperature and baryon chemical potential region in the phase diagram. Measurements from such a beam energy scan of heavy-ion collisions provide us with a unique opportunity to quantitatively study the nature of the QCD phase transition from hadron gas to the QGP at different net baryon densities. 

Figure~\ref{fig:QCD_phase_diagram} presents our current somewhat limited knowledge of the nuclear matter phase diagram. A recent lattice QCD calculation of higher-order susceptibilities at $\mu_B = 0$ allows for a Taylor series extrapolation of the thermodynamic quantities to moderate finite $\mu_B$ \cite{Bazavov:2018mes}. This work showed that the phase transition from HRG to QGP remains to be a smooth cross-over to $\mu_B \sim 250-300$ MeV. The phase diagram region with $\mu_B < 300$ MeV corresponds to mid-rapidity heavy-ion collisions with a collision energy $\sqrt{s} \gtrsim 15$\,GeV. The estimated crossover line agrees with the chemical freeze-out temperatures and net baryon chemical potentials extracted from the STAR BES hadron yield measurements \cite{Andronic:2017pug,Adamczyk:2017iwn}. 
The off-diagonal susceptibility correlations offers additional insights to the chemical freeze-out conditions in relativistic heavy-ion collisions \cite{Bellwied:2018tkc, Bellwied:2019pxh}.
The three blobs in Fig.~\ref{fig:QCD_phase_diagram} indicate the averaged fireball trajectories for typical Au+Au collisions at $\sqrt{s} = 200, 19.6$, and $5$\,GeV. A fireball created in the lower collision energy can probe a larger $\mu_B$ but lower $T$ region of the QCD phase diagram.

To establish definitive links between observables and structures in the phase diagram we need detailed dynamical modeling of all stages of heavy-ion collisions. Precise flow measurements of the hadronic final state, together with phenomenological studies, can elucidate the collective aspects of the baryon-rich QGP and extract the QGP transport properties, such as its viscosity and charge diffusion coefficients. Because relativistic heavy-ion collisions have a complex dynamics going through multiple stages, a fully integrated theoretical framework is required to provide reliable estimates of the dynamical evolution of the collisions and all relevant sources of fluctuations.

Over the past decade, extensive phenomenological studies have been focused on relativistic heavy-ion collisions at LHC and the top RHIC energies (see e.g.\cite{Heinz:2013th, Gale:2013da}, for a review). 
Recently, more and more interests have been shifting towards studying heavy-ion collisions at the intermediate energy regime. At $\sqrt{s} \sim \mathcal{O}(10)$\,GeV, heavy-ion collisions strongly violate longitudinal boost invariance and require full 3D modeling of their dynamics \cite{Noronha:2018atu, Shen:2020jwv}. It is important to employ initial conditions with non-trivial rapidity dependence. The complex 3D collision dynamics can be approximated by parametric energy deposition \cite{Hirano:2005xf, Bozek:2010vz, Bozek:2015bna, Bozek:2017qir, Shen:2020jwv, Sakai:2020pjw}. Non-trivial dynamics could be included to model the energy loss at individual nucleon-nucleon collision impact. Initial state models have been built based on classical string deceleration \cite{Shen:2017bsr, Bialas:2016epd}. And there are 3D initial conditions based on hadronic transport simulations \cite{Pang:2012he, Karpenko:2015xea, Du:2018mpf}.
They provide interesting correlations between the longitudinal energy distribution and flow velocity.
The earlier work of Anishetty, Koehler and McLerran (AKM) in 1980 \cite{Anishetty:1980zp}, which found that nuclei were significantly compressed and excited when they collide at extreme relativistic energies, was pursued only sporadically. This AKM picture was generalized recently to understand early-stage baryon stopping from the Color Glass Condensate-based approaches in the fragmentation region \cite{Li:2018ini, McLerran:2018avb}. The incoming nucleons or valence quarks get decelerated and compressed by the strong shock wave of small-x gluons. Compared to the previous phenomenological approaches, this CGC-based approach is \textit{ab initio} calculations of baryon stopping at high energies. This approach becomes less applicable at intermediate collision energies.
Last but not least, there are recent theory developments from a holographic approach at intermediate couplings to understand the initial energy density and baryon charge distributions \cite{Attems:2018gou}.

\begin{figure}[t!]
    \centering
    \includegraphics[width=1.0\linewidth]{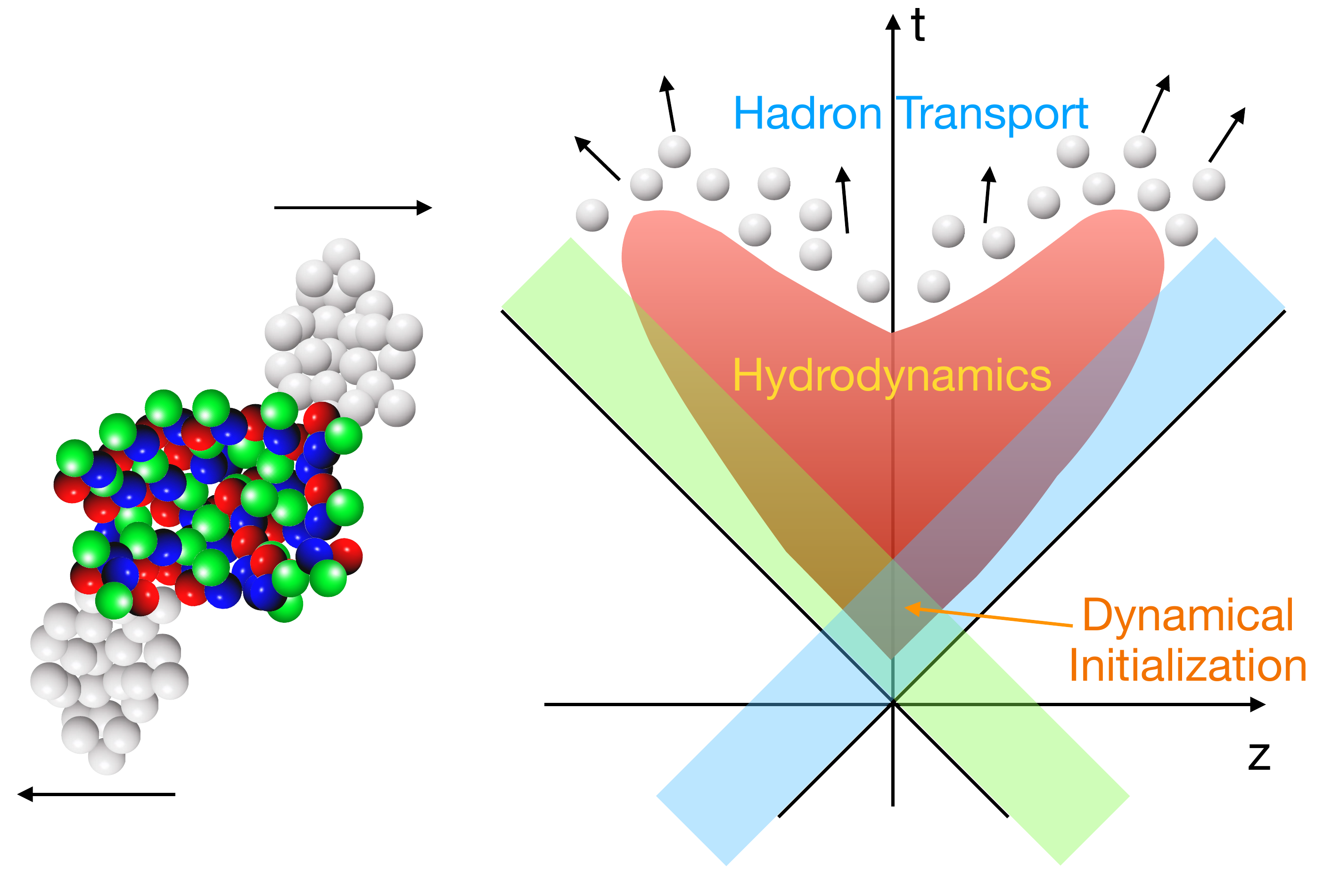}
    \caption{An illustration of relativistic heavy-ion collisions at an intermediate collision energy. When the two colliding nuclei overlap with each other at early time, dynamical initialization connects initial state collision impact with hydrodynamics. }
    \label{fig:3D_collision}
\end{figure}

At a lower the collision energy, the longitudinal Lorentz contraction becomes weaker on the colliding nuclei. The overlapping time for the two nuclei to pass through each other is significant, $\tau_\mathrm{overlap} \sim 2R/\sinh(y_\mathrm{beam})$ \cite{Karpenko:2015xea, Shen:2017bsr, Shen:2017fnn}. Here the $R$ is the nucleus radius and the beam rapidity $y_\mathrm{beam} = \mathrm{arccosh}(\sqrt{s}/(2m_N))$. Therefore, it is important to understand the pre-equilibrium dynamics during this period. Dynamical initialization schemes have been developed to model this extended interaction region in heavy-ion collisions (see Fig.~\ref{fig:3D_collision}). They interweave the initial collision stage with hydrodynamics on a local basis while the two nuclei pass through each other. The initial state energy-momentum and conserved charge density currents are treated as sources input to the hydrodynamic fields at different time, 
\begin{eqnarray}
\partial_\mu T^{\mu\nu} &=& J^\nu_\mathrm{source}(\tau, \vec{x}) \\
\partial_\mu J^\mu_B &=& \rho_{B,\mathrm{source}}(\tau, \vec{x}).
\end{eqnarray}
Such a dynamic initialization scheme was initially proposed in Refs.~\cite{Okai:2017ofp, Shen:2017ruz} and has been adopted by several groups \cite{Shen:2017bsr,Du:2018mpf, Akamatsu:2018olk, Kanakubo:2019ogh}.

Solving the equations of motion of hydrodynamics at low energies requires an equation of state (EoS), which describes the thermodynamic relations of nuclear matter at finite baryon density. The current lattice QCD techniques can not directly compute such an EoS because of the sign problem \cite{Ratti:2018ksb}. However, at vanishing net baryon density or $\mu_B = 0$\,GeV, higher-order susceptibilities have been computed by lattice QCD \cite{Bazavov:2018mes}. These susceptibility coefficients were used to construct nuclear matter EoS at finite baryon densities through a Taylor expansion \cite{Monnai:2019hkn, Noronha-Hostler:2019ayj, Parotto:2018pwx}. These estimated EoS are reliable within the region where $\mu_B/T \lesssim 2$ in the phase diagram. For the region, where temperature is below $\sim 150$ MeV, the lattice QCD EoS is glued with an EoS for hadron resonance gas (HRG). To ensure energy-momentum conservation in the hydrodynamics + hadronic transport approaches, the particle species in the HRG EoS needs to be the same as those in the transport model. A mismatch in the particle content in the HRG EoS could lead to 5-10\% variations in the particle yields and flow observables \cite{Paquet:2020rxl}. At $\mu_B = 0$\,GeV, matching EoS between the two phases was done on the trace anomaly \cite{Huovinen:2009yb, Moreland:2015dvc}. While at finite baryon density, susceptibility coefficients are matched individually and then the thermal pressure was constructed. Full-fledged hydrodynamics + hadronic transport simulations with a EoS at finite baryon density, \textsc{NEoS}, has been applied to heavy-ion collisions at intermediate collision energies \cite{Monnai:2019hkn}. That work found that the enforcement of strangeness neutrality improved the description of relative particle yields for multi-strange particles measured in Pb+Pb collisions at the top SPS energy .

The dynamical initialization and EoS at finite baryon density are two essential ingredients to enable hybrid simulations for heavy-ion collisions at intermediate collision energies. A fully integrated framework \cite{Shen:2017bsr,Shen:2018pty,Monnai:2019hkn} was shown to reproduce the rapidity dependence of particle production as well as the collision energy dependence of the STAR $v_n\{2\}\,(n = 2,3)$ flow measurements in Au+Au collisions from 200 GeV to 7.7 GeV \cite{Adamczyk:2017hdl, Shen:2020jwv}. Remarkably, this preliminary calculation can produce a similar non-monotonic collision energy dependence present in the experimental triangular flow data measured at the RHIC BES phase I \cite{Adamczyk:2016exq}, without the need for a critical point in the phase diagram. Therefore, it is essential to understand the interplay among the duration of dynamical initialization, the variation of the speed of sound, and the $T$- and $\mu_B$-dependence of the specific shear viscosity.
That work demonstrated a critical role of theoretical modeling in elucidating the origin of the non-monotonic behavior seen in the RHIC measurements. Phenomenological studies of the precise anisotropic flow measurements from the upcoming analysis of RHIC BES II will be able to further constrain the $\mu_B$ dependence of the QGP shear and bulk viscosities \cite{Karpenko:2015xea, Shen:2020jwv}.

Modeling relativistic heavy-ion collisions beyond the Bjorken boost invariant offers us a new dimension to study the fluctuations of flow anisotropy as a function particle rapidity.
The flow correlations between different rapidity regions reflect the longitudinal fluctuations of local energy density profiles in the full 3D dynamics \cite{Bozek:2017qir, Li:2019eni, Franco:2019ihq, McDonald:2020oyf}. The fluctuations of the anisotropic flow coefficients along the longitudinal direction elucidates how the shape of fireball varies as a function of rapidity. The longitudinal fluctuations will cause the initial eccentricity vectors to fluctuate from one space-time rapidity to another.
As a consequence, the anisotropic flow coefficients decorrelate as a function of particle rapidity. A recent systematic study \cite{Sakai:2020pjw} showed that initial state fluctuations and thermal fluctuations in hydrodynamics were equally important to understand the centrality dependence of flow decorrelation measurements at LHC. The RHIC beam energy scan program will systematically study the collision energy dependence of the rapidity flow fluctuations from 200 GeV down to 7.7 GeV. The RHIC BES program offers a unique opportunity to study interplay between thermal production and collision transport (stopping mechanisms). Such measurements combined with phenomenological modeling can lead to strong constraining power for our understanding of the longitudinal dynamics in heavy-ion collisions \cite{Shen:2017bsr}. Future measurements with identified particles have potentials to further shed lights on initial distributions of conserved charges (net baryons, strangeness, and electric charges) at different collision energies. 

The realistic dynamical simulations of heavy-ion collisions at the RHIC BES energies pave the foundation to quantitatively understand the out-of-equilibrium stochastic fluctuations when the collision systems evolve close to a conjectured QCD critical point in the phase diagram. Near the QCD critical point, the relaxation times of the critical fluctuations grow large and rapidly become comparable with the system size \cite{Stephanov:2004wx, Bzdak:2019pkr}. The rapid expansion of the collision system drives the fluctuations related to the QCD critical point out-of-equilibrium \cite{Stephanov:2017ghc, Akamatsu:2018vjr, An:2019csj}. Therefore, these ``critical slowing down'' dynamics require realistic event-by-event simulations to address how far they are out-of-equilibrium quantitatively.

There are two primary approaches to this problem in the literature. One is to explicitly evolve these fluctuations in a stochastic hydrodynamics framework \cite{lifshitz2013statistical}. In heavy-ion physics, numerical simulations of stochastic hydrodynamics were developed by several groups to study thermal and critical fluctuations \cite{Young:2014pka, Singh:2018dpk, Bluhm:2020mpc, De:2020yyx, Nahrgang:2020yxm}. However, because these stochastic fluctuations are local $\delta$-functions in the coordinate space, one needs to take care numerical cut-off dependence to ensure that the simulation results are physical. The other approach studies the deterministic evolution of the two-point correlation function of the fluctuations. This approach was pioneered by Andreev in the 1970’s in the nonrelativistic case \cite{andreev1978corrections}. This approach is often referred to as ``hydro-kinetic'' \cite{Akamatsu:2016llw, Stephanov:2017ghc, Akamatsu:2018vjr, Martinez:2018wia, An:2019osr, An:2019csj}.

The hydro-kinetic approach should be consistent with the stochastic hydrodynamics approach for the two-point correlation function. A side-by-side comparison between these two approaches will be extremely valuable to improve both theories. On the one hand, the renormalization of the hydrodynamic equations are well under control in the hydro-kinetic framework. It can guide the stochastic hydrodynamics on how to introduce a UV cut-off to regulate multiplicative noise in the simulations. On the other hand, stochastic hydrodynamics can estimate how much non-linearity in the fluid dynamics will affect the evolution of two-point function, which is neglected in the hydro-kinetic formalism. In the meantime, it is straight-forward to access higher-order correlation functions in the stochastic hydrodynamics.

There are recent studies of the deterministic ``hydro-kinetic'' formalism on simplified (1+1)D hydrodynamic backgrounds \cite{Rajagopal:2019xwg, Du:2020bxp}. Those work found that the feedback contributions to the thermodynamic quantities from the out-of-equilibrium fluctuations is on the order of $10^{-4}$, which can be safely neglected in the simulations.

To deliver quantitative predictions for experimental signals of the critical fluctuations at the RHIC BES II, we need to further develop the theoretical frameworks in the following directions. In the ``hydro-kinetic'' approach, the effects of flow gradients on the critical fluctuations need to be addressed quantitatively with realistic 3D event-by-event hydrodynamic simulations. Although the works~\cite{An:2019osr, An:2019csj} derived the equations of motion for two-point correlation functions under general flow background, there are substantial challenges to implement these equations in the state-of-the-art hydrodynamic framework. In addition, a theoretical formulation is needed to map these two-point correlation functions from the coordinate space to momentum correlations among particles, an essential step to providing theory predictions for measurements. Finally, generalization to $n$-point correlation functions $(n > 2)$ could enhance the sensitivity of the experimental signals, but requires substantial efforts in the theoretical side \cite{Pratt:2019fbj}.

\subsection{Quantitative characterization of the QGP transport properties}

\begin{figure*}[t!]
    \centering
    \includegraphics[width=1.0\linewidth]{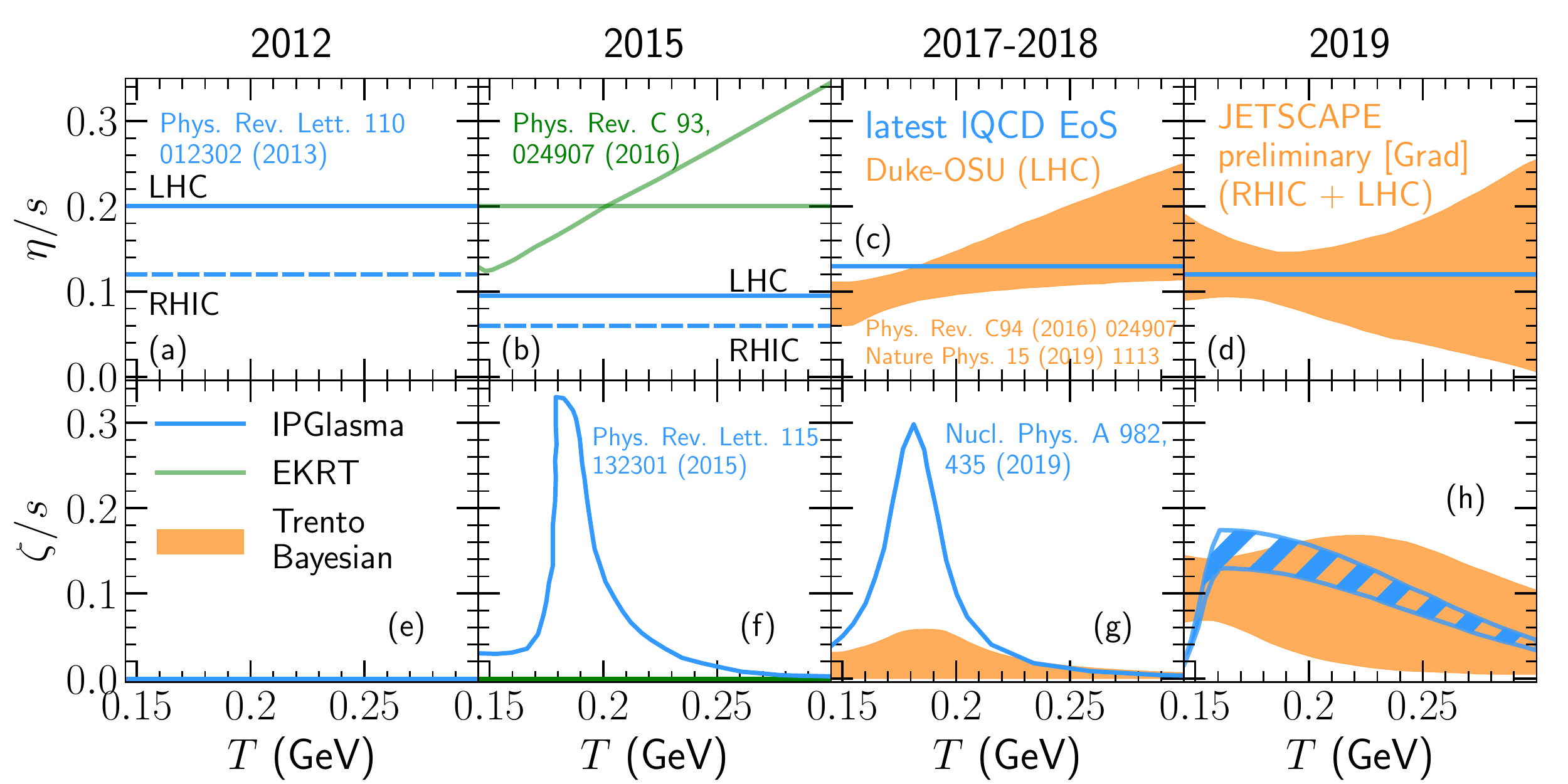}
    \caption{A summary of phenomenological constraints on the temperature-dependent QGP specific shear and bulk viscosities from 2012 to 2019. This figure is taken from \cite{Shen:2020gef}. The QGP viscosities are constrained by the collective flow measurements at the top RHIC and LHC collision energies. Results from open-source dynamical frameworks using three different initialization models are compared among each others. The IP-Glasma initial conditions with hydrodynamics + hadronic transport approach continuously improving the constraints on the QGP viscosity \cite{Gale:2012rq, Ryu:2015vwa, Schenke:2018fci, Schenke:2020unx, Schenke:2020mbo, Gale:2020xlg}. The blue hatched area in (h) indicates the variation of $(\zeta/s)(T)$ with (upper) and without (lower) a pre-equilibrium K{\o}MP{\o}ST effective kinetic theory (EKT) phase \cite{Gale:2020xlg}. The result using the EKRT initial condition is from Ref. \cite{Niemi:2015qia}. The Bayesian extracted QGP $(\eta/s)(T)$ and $(\zeta/s)(T)$ \cite{Bernhard:2016tnd, Bernhard:2019bmu} was constrained by flow measurements in p+Pb and Pb+Pb collisions at the LHC. The orange bands indicate a 90\% confidence level. The preliminary Bayesian analysis calibrated with combined RHIC and LHC flow measurements was presented by the JETSCAPE Collaboration \cite{Paquet:2020rxl, Everett:2020yty}.}
    \label{fig:viscosity_summary}
\end{figure*}

\subsubsection{Charge diffusion}

The fluid dynamics of the bulk QGP medium is driven by the local pressure gradients. The RHIC BES program allows us to study the evolution of non-vanishing conserved charge currents, which is additionally dependent on the gradients of chemical potentials. Therefore, the interplay among different gradient forces in the dynamics of conserved charge currents can elucidate novel transport processes inside the medium, namely the charge diffusion constants and heat conductivity. These QGP transport coefficients are so far poorly constrained but equally important as the specific shear and bulk viscosity. Because individual quark carries multiple quantum charges, the diffusion currents of conserved charges are coupled with each other. Quantitative understanding how multiple conserved charge currents diffuse in and out of the expanding fluid cells has been leading the development of the next generation of dynamical framework. Such a framework can unravel the detailed chemical aspects of the QGP.
This topic has been stimulating interests in developing realistic initial conditions for conserved charge distributions \cite{Shen:2017bsr, Du:2018mpf, Martinez:2019jbu, Martinez:2019rlp} and detailed modeling of the QGP chemistry \cite{Pratt:2012dz, Pratt:2018ebf, Pratt:2019pnd, Monnai:2019hkn}.

The net baryon diffusion is driven by the local gradients of net baryon chemical potential over temperature in the hydrodynamic evolution. Causal Israel-Stewart-like equations of motion for the net baryon diffusion current were derived based on the Grad's 14-moment and Chapman-Enskog methods \cite{Monnai:2012jc, Denicol:2012cn, Jaiswal:2015mxa}. Recently, such formulations were generalized to include multiple diffusion currents together with their cross couplings \cite{Fotakis:2019nbq}. The functional forms of the transport coefficients for the diffusion matrix have been studied in the transport models \cite{Greif:2017byw, Rose:2020sjv}. Additional coupling terms with the shear and bulk viscous tensors appear at the second order in the gradient expansion \cite{Denicol:2012cn, Jaiswal:2015mxa}.

Phenomenologically, the net baryon diffusion current transports more baryons from forward rapidities to the mid-rapidity region \cite{Monnai:2012jc, Denicol:2018wdp, Li:2018fow, Du:2019obx}. The gradients of $\mu_B/T$ act against local pressure gradients and decelerate baryon charges relative to the bulk fluid cells along the longitudinal direction. Therefore, the shape of the rapidity distribution of the net protons show a strong sensitivity to the baryon diffusion \cite{Denicol:2018wdp, Li:2018fow}. The cross diffusion between the net baryon and net strangeness induces an oscillating distribution of the net strangeness current at late time \cite{Fotakis:2019nbq}. It will be interesting to see how this pattern is mapped to final state hadron correlations, such as Kaons and $\Lambda$s. Measurements of identified particle rapidity distributions will play an important role to unravel the charge diffusion processes in heavy-ion collisions. 

The net baryon diffusion process in the hydrodynamic phase can only transport the net baryon charges by $\sim$1 unit in rapidity \cite{Denicol:2018wdp}. As the bulk fluid is rapidly exploding along the $z$ direction, transporting net baryon charge back to the mid-rapidity during the hydrodynamic evolution is very challenging at high energies. It turned out to be difficult to reproduce the small but non-zero net proton rapidity distribution at 200 GeV measured by the BRAMHS Collaboration \cite{Li:2018fow} only through baryon diffusion. 
The measurement suggests there is a large baryon stopping at early stage of the heavy-ion collisions. Allowing the baryon charge to fluctuate to the string junction \cite{Kharzeev:1996sq} in the initial state, we can reproduce the net baryon rapidity distributions at 200 GeV. In fact, this model can consistently reproduce the net proton distribution measured by STAR Collaboration down to 7.7 GeV at mid-rapidity~\cite{Shen:2020jwv}. The rapidity distribution in the RHIC BES phase II will further help us to constrain the initial state baryon stopping in this phenomenological model. 
Because the net proton rapidity distribution is sensitive to both the initial state stopping and baryon diffusion \cite{Li:2018fow}, independent experimental observables are needed to disentangle these two effects. 

Recently, charge balance functions of identified hadrons were proposed as independent observables to constrain the charge diffusion constants of the QGP \cite{Kapusta:2017hfi, Pratt:2019pnd, De:2020yyx}. Quark-anti-quark pairs can be thermally produced in local fluid cells. As a $q\bar{q}$ pair randomly walks through the medium, it will develop a finite correlation length which is controlled by the local fluid expansion and charge diffusion constants. This spatial correlation eventually maps to momentum correlations among hadrons which the $q\bar{q}$ pair hadronizes into. 
In Ref.~\cite{Pratt:2019pnd}, the authors found that a larger diffusion constant in the QGP medium leads to wider azimuthal and rapidity distributions for the $K^+K^-$ and $p\bar{p}$ correlations. In the meantime, the underlying hydrodynamic flow acts a boost to these correlation functions, which results in a focusing of the azimuthal distributions of balance functions. Therefore, measurements of balance functions can provide an independent constraint on the QGP charge diffusion constants.

\subsubsection{Specific shear and bulk viscosities}
\label{Sec:viscosity}

The shear and bulk viscosities, normalized by entropy density, characterize the dissipation of energy-momentum currents in a medium. Extracting the specific QGP viscosity has been the main theme of phenomenological studies in relativistic heavy-ion collisions over the past 20 years. Because of the strongly-coupled nature of QGP, it is difficult to calculate these transport coefficients from first principles. In the meantime, phenomenological constraints drawn from comparing with the precise anisotropic flow measurements have been leading our field to understand the properties of the QGP. To systematically constrain the QGP transport properties, adopting the Bayesian statistical analysis has become a standard approach in our field \cite{Novak:2013bqa, Pratt:2015zsa}. 

Figure~\ref{fig:viscosity_summary} summarizes the collective effort of quantifying the QGP transport properties over the past eight years in our field. As theoretical tools are being developed rapidly to include more realistic physics, the extraction of QGP transport properties becomes more systematic. Chronologically, before the implementations of bulk viscous effects in the dynamical models, the saturation-based IP-Glasma and EKRT initial conditions prefer an effective shear viscosity of $0.12-0.20$ in the hydrodynamic phase to achieve a simultaneous description of all orders of harmonic flow coefficients \cite{Gale:2012rq, Niemi:2015qia}. Owing to large pressure gradients and finite initial radial flow in the IP-Glasma initial conditions, a temperature-dependent bulk viscosity is essential to balance the consequent strong flow and reproduce the mean $p_T$ measurements in the heavy-ion collisions. At the same time, the introduction of the bulk viscosity reduced the extracted QGP shear viscosity by almost 50\% \cite{Ryu:2015vwa, Ryu:2017qzn}. By adopting the equation of state from the latest lattice QCD calculations \cite{Borsanyi:2013bia, Bazavov:2014pvz}, the Duke-OSU group delivered the first Bayesian Inference on the temperature-dependent shear and bulk viscosities \cite{Bernhard:2016tnd, Bernhard:2019bmu} using the flow measurements at LHC. There was tension of the Bayesian extracted bulk viscosity $(\zeta/s)(T)$ \cite{Bernhard:2016tnd,Bernhard:2019bmu} with the parameterization used in the IP-Glasma hybrid framework \cite{Schenke:2018fci} in 2018. This difference was greatly reduced in the 2019 updated simulations. On the one hand, the significant changes in the IP-Glasma hybrid framework come from allowing the peak temperature of bulk viscosity to drop from $T_\mathrm{peak} = 180$ MeV to $160$ MeV \cite{Schenke:2020unx, Schenke:2020mbo}. A lower $T_\mathrm{peak}$ with a smaller $(\zeta/s)_\mathrm{max}$ is favored by hadron mean $p_T$ measurements in peripheral Pb+Pb and Au+Au collisions, in which the maximum temperatures at the starting time of hydrodynamic simulations are close to, or even below, $180$ MeV. On the other hand, a more flexible prior parameterization of $(\zeta/s)(T)$ in the JETSCAPE preliminary Bayesian analysis allows a large QGP bulk viscosity in the posterior distribution for the model to reproduce flow measurements \cite{Paquet:2020rxl, Everett:2020yty}.

The preliminary Bayesian analysis from the JETSCAPE Collaboration demonstrated that the first simultaneous calibration using flow observables at the top RHIC and LHC energies, which differ by an order of magnitude. Such combined analysis showed strong constraints on the temperature dependence of the QGP shear and bulk viscosities around the cross-over temperature region \cite{Paquet:2020rxl, Everett:2020yty}. Heavy-ion collisions at RHIC and the LHC offer a wide dynamical range for the theoretical framework to explore the parameter space. At the same time, different types of off-equilibrium corrections at the particlization stage introduce sizable theoretical uncertainties to the Bayesian extraction, demanding more theoretical work in this direction. That work also emphasized that performing closure tests in Bayesian analysis was an essential step before extracting any physical information from experimental measurements \cite{Paquet:2020rxl, Everett:2020yty}.

The next generation Bayesian analysis requires improving the start-of-the-art theoretical framework to access new physics in the dynamical simulations. A recent work bridges the event-by-event IP-Glasma initial state \cite{Schenke:2012wb, Schenke:2020mbo} and viscous hydrodynamics \cite{Schenke:2010rr, Paquet:2015lta} using an effective QCD kinetic theory (EKT), K{\o}MP{\o}ST \cite{Kurkela:2018wud, Kurkela:2018vqr}. The EKT in the pre-equilibrium stage can drive the collision system sufficiently near local thermal equilibrium and smoothly matches the full energy-momentum tensor to viscous hydrodynamics.
Remarkably, this framework can quantitatively reproduce a variety of flow measurements in heavy-ion collisions from 200 GeV to 5020 GeV \cite{Gale:2020xlg} with an effective $(\eta/s)_\mathrm{eff} = 0.12$ and a temperature-dependent bulk viscosity (see Fig.~\ref{fig:viscosity_summary}h). A direct comparison with the simulations without the EKT phase \cite{Schenke:2020mbo} showed that the conformal EKT generates a faster expansion than viscous hydrodynamics at this early stage. A similar finding was earlier observed for free-streaming dynamics \cite{Liu:2015nwa}. This additional pre-equilibrium phase thus leads to a 35\% larger extracted QGP bulk viscosity to describe identified particle mean $p_T$. A recent work \cite{NunesdaSilva:2020bfs} pointed out that the breaking of conformality in the pre-equilibrium stage could be important in the extraction of bulk viscosity. Those studies demonstrated the significant phenomenological impact of a realistic modeling of the early stage of heavy-ion collisions on constraining the QCD bulk viscosity.

\subsection{Challenges and opportunities in small systems}

The RHIC and the LHC collide a variety types of nuclei, which offer us measurements to study the collectivity as a function of the collision system size. As the size of a QGP droplet shrinks, the lifetime for the strongly-coupled hydrodynamic evolution becomes shorter and shorter. Therefore, the final state particles' momentum distributions and correlations can reveal more information about the early-stage dynamics \cite{Romatschke:2015gxa}. On the theoretical side, small systems are the phenomenological ground to understand how heavy-ion collisions achieve macroscopic hydrodynamic behavior, chemical, and kinetic equilibration from far-out-of-equilibrium. Non-hydrodynamic modes can potentially play an important role and push the hydrodynamic framework to its limits. At the same time, the increasing roles of fluctuations and non-flow correlations stress the ability of experiments to unambiguously identify flow signatures.

\begin{figure*}[ht!]
    \centering
    \includegraphics[width=1.0\textwidth]{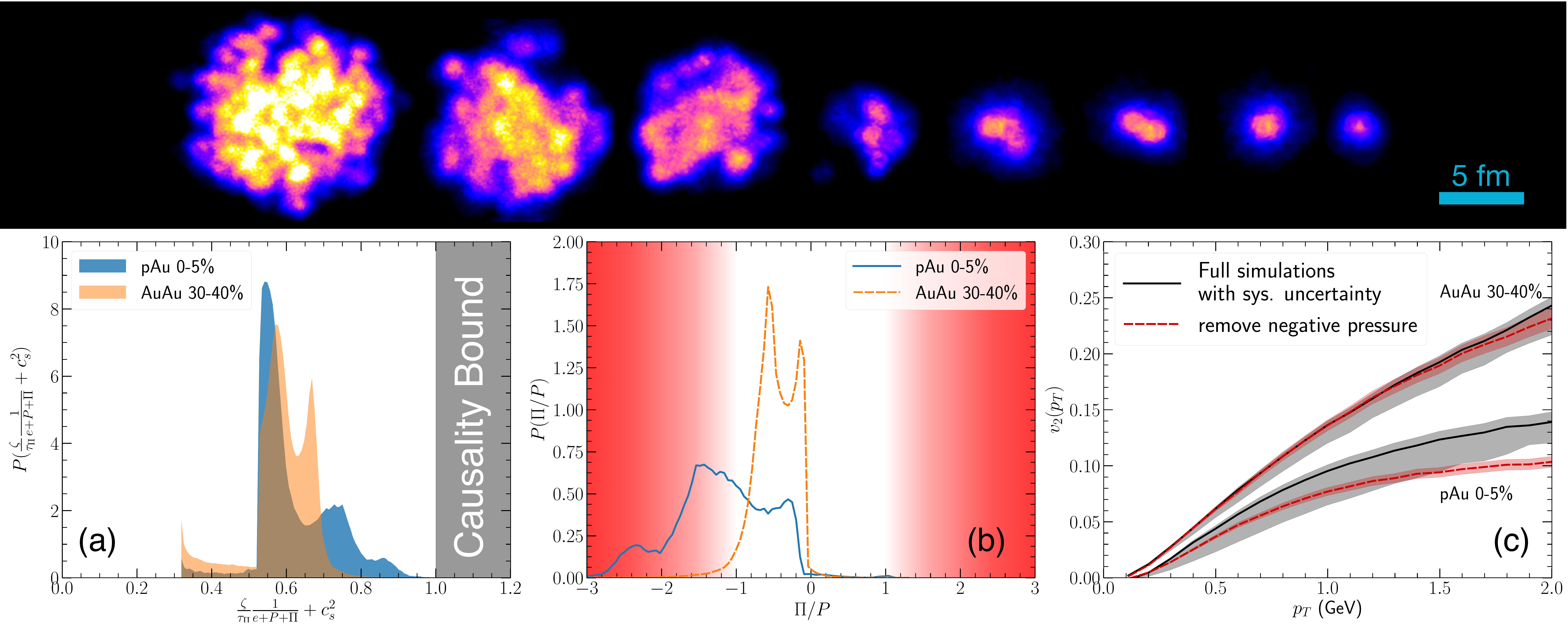}
    \caption{Upper panel: The color contour plots for the initial event-by-event energy densities of U+U, Au+Au, Ru+Ru, O+O, $^3$He+Au, d+Au, p+Au, and p+p collisions at 200 GeV from IP-Glasma initial conditions \cite{Schenke:2020mbo}. Panel (a): Check the non-linear causality condition with bulk viscous pressure \cite{Bemfica:2019cop} in a typical 0-5\% p+Au and 30-40\% Au+Au collisions at 200 GeV. Panel (b): The distribution of the ratio of bulk viscous pressure over thermal pressure, $\Pi/P$ in fluid cells for 0-5\% p+Au and 30-40\% Au+Au collisions. Panel (d): The effect of regulating negative total pressure on the elliptic flow coefficients in 0-5\% p+Au and 30-40\% Au+Au collisions at 200 GeV \cite{Schenke:2019pmk}.}
    \label{fig:small_system_challenges}
\end{figure*}

The collective flow and its hydrodynamic description is most robust in central Pb+Pb and Au+Au collisions. Therefore, by first constraining model parameters with flow measurements in heavy-ion collisions and then extrapolating to small systems, we can provide a stringent parameter-free test for hydrodynamic models. Such a study was pioneered in Ref.~\cite{Weller:2017tsr}. A more systematic study has recently been carried out using the state-of-the-art IP-Glasma + MUSIC + UrQMD hybrid framework  \cite{Schenke:2020mbo}. That work presented a remarkable success in describing the system size dependence of the flow measurements over more than two orders of magnitude in particle multiplicity. With a single set of model parameters, this theoretical framework can quantitatively describe particle production, radial and anisotropic flow observables, and multi-particle correlations from Pb+Pb to p+Pb collisions at the LHC and from Au+Au and to p+Au collisions at the top RHIC energy. That work demonstrated that the universal hydrodynamic response to collision geometry dominated the flow production in collisions with $dN^\mathrm{ch}/d\eta \ge 10$ at mid-rapidity. On the other hand, the correlation between $v_2$ and initial momentum anisotropy from the pre-hydrodynamic phase becomes stronger in the lower multiplicity collisions \cite{Schenke:2019pmk}. That work provided phenomenological evidence that the flow in low multiplicity collisions can elucidate the early dynamics of the collisions. A similar finding was shown from a study of the correlation between the system's elliptic flow coefficient and initial energy-momentum tensor \cite{Sousa:2020cwo}.

The anisotropic flow in high energy p+p collisions still challenges our understanding of the underlying dynamics in these small systems. The computed two-particle cumulant $v_2\{2\}$ from the IP-Glasma hybrid framework increases as charged hadron multiplicity decreases \cite{Schenke:2020mbo}, which is not seen in the flow measurements at the LHC. Comparing with results from \cite{Habich:2015rtj, Weller:2017tsr} suggests that the momentum anisotropy from the pre-equilibrium Glasma phase might be too strong in p+p collisions. 
Moreover, recent work pointed out that pure hydrodynamic evolution introduces a too strong non-linear cubic response of elliptic flow to the initial eccentricity in p+p collisions, which results in a positive four-particle cumulant $C_2\{4\}$ opposite to the experimental measurements \cite{Zhao:2020pty}. Elucidating the real dynamics in p+p collisions requires more theoretical progress on quantifying the contributions from pre-equilibrium dynamics, thermal and longitudinal fluctuations, non-trivial correlations from local conservation laws, and non-hydrodynamic modes.

Finally, it is instructive to quantify to what extent small collision systems have been pushing the hydrodynamic framework to its limits. Conditions for non-linear causality bounds of second-order hydrodynamics were derived for radially expanding systems \cite{Floerchinger:2017cii} and for a general flow background \cite{Bemfica:2019cop, Bemfica:2020xym}. These non-linear causality bounds set strong constraints on the maximal allowed viscous pressure in dynamical simulations. One can check the non-linear causality condition including the bulk viscous pressure by computing the following ratio,
\begin{equation}
	r = \frac{\zeta}{\tau_\Pi} \frac{1}{e + P + \Pi} + c_s^2.
\end{equation}
Causality condition requires the ratio $r < 1$ \cite{Bemfica:2019cop}.
Fig.~\ref{fig:small_system_challenges}a compares the distribution of this ratio from individual fluid cell in a typical 0-5\% p+Au collision with that from a 30-40\% Au+Au collisions.
Because of the strong expansion rate and the consequent large negative bulk viscous pressure, the causality conditions in a typical p+A collision are $\sim$20\% closer to the bound than those in an A+A collision.

The large pressure gradients and the consequent violent expansion in small systems can result in negative total (thermal + bulk viscous) pressure in a significant fraction of fluid cells, shown in Fig.~\ref{fig:small_system_challenges}b. These bubbles may cause unstable cavitation inside the QGP \cite{Rajagopal:2009yw, Habich:2014tpa, Byres:2019xld}. Preferably, one would switch the fluid dynamic description to a dilute transport approach before the total pressure becomes negative. 
In the absence of such an advanced theoretical framework, we can roughly estimate the phenomenological impact of these negative pressure regions on final flow observables by numerically regulating the size of bulk viscous pressure to be less than the thermal pressure. This modification leads to a sizable variation in the elliptic flow in 0-5\% p+Au collisions in Fig.~\ref{fig:small_system_challenges}c. For $p_T < 1.5$ GeV, this associated theoretical uncertainty is comparable to effects resulting from varying the second-order transport coefficients \cite{Schenke:2019pmk}. The situation in semi-peripheral A+A collisions is much better with a negligible effect on final flow observables. While this ad hoc numerical regulation can only provide us with a rough estimate, Fig.~\ref{fig:small_system_challenges}c demonstrates that the phenomenological descriptions of small systems with the standard second-order viscous hydrodynamics are approaching to the limits of the model. 
Anisotropic hydrodynamics, which reproduces the free-streaming limit at large viscosity, is a good theoretical tool to provide us with more robust guidance on this issue in small systems \cite{Alqahtani:2017tnq, McNelis:2018jho}. 

\subsection{Interdisciplinary connections}

Relativistic heavy-ion collisions embrace a richness of physics that expands multiple energy scales. At the early stages of the collisions, the dynamics of approaching hydrodynamic behavior from far-out-of-equilibrium has strong connections with reheating of the early universe following inflation and over-occupied cold atomic gases (see a recent review \cite{Berges:2020fwq}). The collision geometry in heavy-ion collisions can be used as a powerful microscope to image the structures inside nuclei and nucleons. In the meantime, it has become an active topic to apply cutting-edge machine learning techniques to study the complex dynamics of heavy-ion collisions.

\subsubsection{Nuclear structure physics}

Although the bulk evolution in relativistic heavy-ion collisions is above 100 MeV in temperature, nuclear structure physics at a much lower energy scale still can play an important role in precision studies of flow observables. 
The Monte-Carlo Glauber model is a key element in computing the initial states for hydrodynamical modeling of ultrarelativistic heavy-ion collisions. The spatial configurations of nucleons inside the colliding nuclei are inputs based on nuclear structures physics for each collision event.
Hydrodynamics can then efficiently transform the shape of the initial energy density profile to the momentum anisotropy of final state particles. Therefore, the measurements of anisotropic flow coefficients and their fluctuations provide us with a tool to \textit{image} the event-by-event shape fluctuations of the colliding nuclei.
Furthermore, small collision systems, such as proton-lead collisions, can image the sub-nucleon fluctuations inside protons in a similar fashion \cite{Schenke:2014zha, Mantysaari:2017cni, Mantysaari:2020axf}.

The structure of nuclei has been approximated for a long time by independent particle models in which the nucleons inside a nucleus are treated as a collection of free and point-like fermions. The colliding nuclei in the relativistic heavy-ion collisions are highly Lorentz boosted along the beam direction. In the lab frame, because of the relativistic time dilation effect, individual nucleons are frozen in their spatial positions as the two nuclei collide with each other. The spatial positions of the nucleons inside the nucleus are usually sampled independently from parametric Woods-Saxon distributions. More realistically, correlations among nucleons at short distance are not negligible, known as short-range correlations (SRC). These nucleon-nucleon (NN) correlations were unambiguously observed in a series of dedicated experiments \cite{Shneor:2007tu, Hen:2016kwk, Duer:2018sxh, Schmidt:2020kcl}. Studies in Ref.~\cite{Alvioli:2009ab, Broniowski:2010jd, Alvioli:2011sk} showed that including realistic NN correlations had sizable effects on the generated initial eccentricity of the energy profile.
The NN correlations have sizable effects on the spectrum of anisotropic flow coefficients in central heavy-ion collisions, in which the geometric distortion of the overlapping region from the impact parameter is minimized.
The ratios of the elliptic flow to triangular flow were found to be sensitive to the spatial configurations of the colliding nuclei with and without SRC \cite{Denicol:2014ywa, Shen:2015qta}. A recent work explored the effect of a possible octupole deformation of $^{208}$Pb on the $v_2/v_3$ ratio \cite{Carzon:2020xwp}.

Furthermore, collisions with deformed nuclei are particularly interesting as one can use the nucleus' intrinsic deformation as an additional control to study the hydrodynamic conversion from spatial eccentricity to momentum anisotropy and inform the transport properties of the QGP. At the top RHIC energy, the deformed U+U collisions were studied together with Au+Au collisions. Full-overlap U+U collisions had the potential to study the hydrodynamic behavior of elliptic flow in the large and dense collision systems while the non-linear path-length dependence of radiative parton energy loss \cite{Heinz:2004ir}. Experimentally, such a study requires to select ``tip-tip'' collisions, defined to occur when the major axes of the Uranium nuclei lie parallel with the beam direction, from the ``body-body'' events, where the major axis of each nucleus is perpendicular to the beam direction \cite{Voloshin:2010ut, Rybczynski:2012av, Goldschmidt:2015kpa}. However, the theoretically proposed triggers based on a two-component Glauber model were not effective in the STAR measurements \cite{Adamczyk:2015obl}. The measurements showed more consistency with the saturation-based initial condition models \cite{Moreland:2014oya, Schenke:2014tga}. Recently, new types of correlations between elliptic flow and mean transverse momentum were proposed, which is sensitive to the nucleus deformation \cite{Giacalone:2019pca}. Similarly, at the LHC energies, $^{129}$Xe+$^{129}$Xe collisions were measured in additional to the $^{208}$Pb+$^{208}$Pb collisions. In contrast to the spherical $^{208}$Pb, the $^{129}$Xe nucleus has a prolate deformation. Hydrodynamic simulations \cite{Giacalone:2017dud} predicted that the elliptic flow coefficients would be 25\% larger in $^{129}$Xe+$^{129}$Xe than those in $^{208}$Pb+$^{208}$Pb in the 0-5\% central collisions. This strong enhancement of $v_2\{2\}$ in central $^{129}$Xe+$^{129}$Xe was confirmed by the LHC measurements \cite{Acharya:2018ihu, Sirunyan:2019wqp, Aad:2019xmh}. 

Stable heavy nuclei are neutron-rich. The neutron density profiles usually have larger RMS radii than those of proton densities, which is referred to as the neutron skin of the nucleus. Measurements of neutron skins in heavy nuclei are critical in studying the equation of state of neutron-rich nuclear matter \cite{Li:2008gp}. In relativistic heavy-ion collisions, neutron skin has a negligible effect on regular flow observables. However, the RHIC proposed to use isobar collisions to minimize the background flow signals in search of Chiral Magnetic Effects (CME). The success of this experimental program relies on a precision level understanding of the flow background in $^{96}_{44}$Ru+$^{96}_{44}$Ru and $^{96}_{40}$Zr+$^{96}_{40}$Zr collisions. A recent work \cite{Hammelmann:2019vwd} pointed out that the neutron skin in $^{96}_{40}$Zr can result in a stronger magnetic field in peripheral collisions and consequently lead to a factor of 2 reductions in the difference in CME signals from $^{96}_{44}$Ru+$^{96}_{44}$Ru vs. $^{96}_{40}$Zr+$^{96}_{40}$Zr collisions. Moreover, realistic nucleon configurations based on density functional theory calculations for $^{96}_{44}$Ru and $^{96}_{40}$Zr nuclei showed sizable differences of the particle elliptic flow $v_2$ between the two isobar collision systems \cite{Xu:2017zcn}.

\subsubsection{Applications of statistics and machine-learning techniques}

\begin{figure}[t!]
    \centering
    \includegraphics[width=1.0\linewidth]{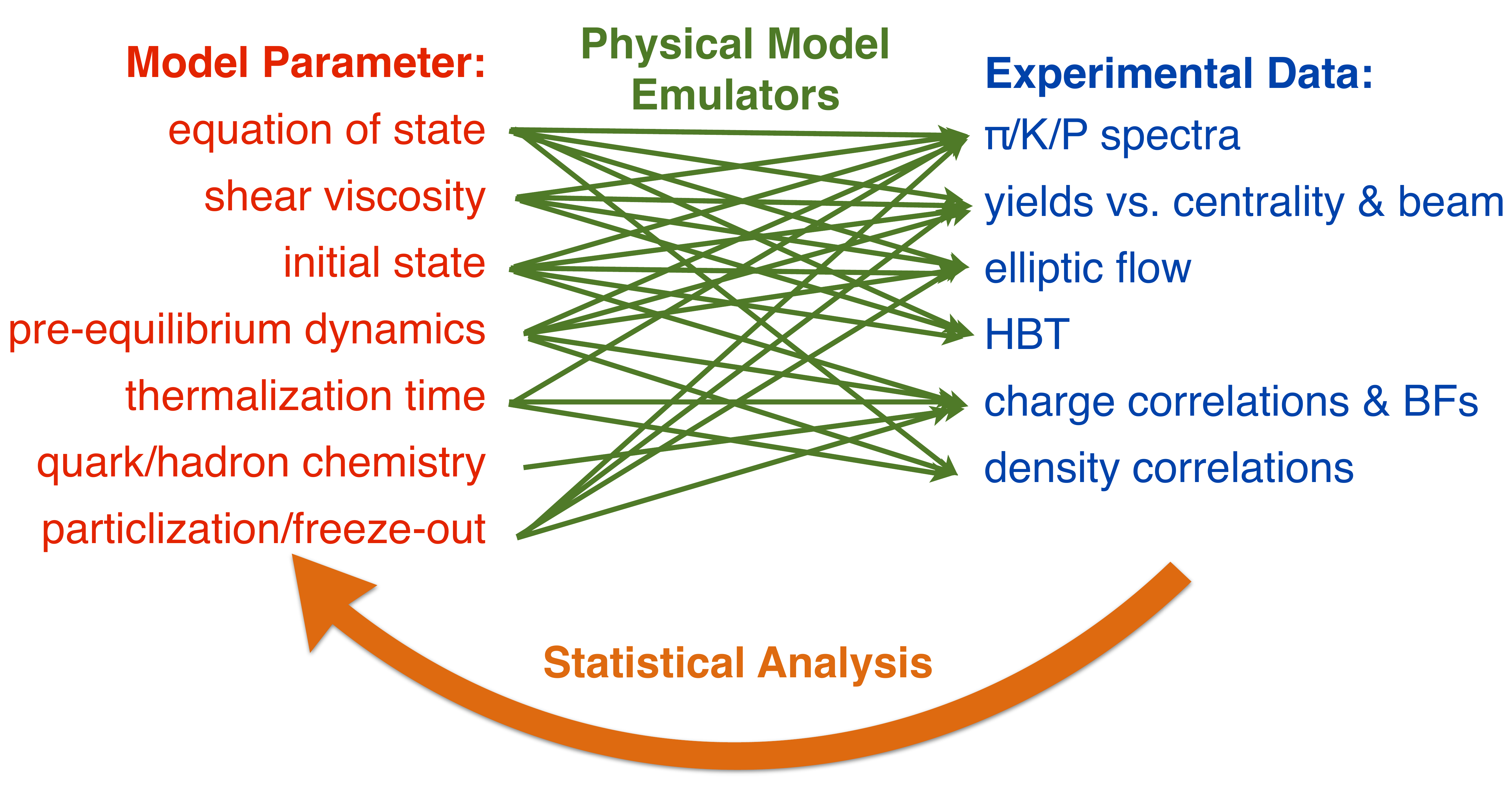}
    \caption{An illustration of the workflow of statistical analysis, such as Bayesian inference, in relativistic heavy-ion collisions. The individual physical parameter goes through the physical model can influence multiple experimental observables. Statistical analysis provides a systematic way to constrain multiple model parameters using a collection of measurements. Trained model emulators, together with Monte-Carlo Markov Chain processes, are often applied in the statistical analysis to explore the high-dimension parameter space efficiently. This figure is modified based on Steffen Bass's talk at Quark Matter 2017 \cite{Bass:QM2017}.}
    \label{fig:Bayesian}
\end{figure}

Extracting quantitative physics from the complex dynamics in relativistic heavy-ion collisions requires producing and analyzing a high-volume of data both in experiments and in numerical simulations. The data acquisition rates at RHIC and LHC experiments grow factorially with the detectors' updates. The precise measurements, such as 6- and 8-particle azimuthal correlations \cite{Khachatryan:2015waa}, are pushing large-scale and high-performance numerical simulations. Parallelization using Graphics Processing Units (GPU) is employed to speed up the event-by-event simulations \cite{Bazow:2016yra, Pang:2018zzo}. 
In the meantime, we need to adopt advanced statistical tools, such as unsupervised learning, Bayesian inference, and deep learning techniques, to systematically analyze the high-volume of simulation data to extract physical properties of the QGP. As shown in Fig.~\ref{fig:Bayesian}, the dynamical modeling of heavy-ion collisions involves multiple model parameters, each of which can influence several experimental measurements. Therefore, it is difficult to identify a single experimental observable to constrain one physical parameter.  
Machine learning is a collection of generic algorithms that allow computers to find non-trivial correlations and patterns in data samples with minimum bias. 

\textbf{Principle component analysis} (PCA) has been widely used to study the roles of fluctuations and correlations in relativistic heavy-ion collisions. PCA is a statistical technique for extracting the dominant components in fluctuating data by transforming a set of correlated variables into independent ones via orthogonal transformations. This method was first introduced to analyze the event-by-event fluctuations of anisotropic flow coefficients \cite{Bhalerao:2014mua, Mazeliauskas:2015vea, Sirunyan:2017gyb} and the breaking of flow factorization ratios \cite{Mazeliauskas:2015efa}. The PCA procedure in that work found two dominant contributions to the two-particle correlation function. The leading component was identified with the event plane anisotropic flow $v_n(p_T)$, while the subleading component was responsible for factorization breaking in hydrodynamics \cite{Mazeliauskas:2015efa}. A recent study \cite{Hippert:2020kde} showed that the subleading principal components of anisotropic flow can reveal details of the hydrodynamic response to small-scale structures in the initial density profiles. Similar studies have also been performed to understand the event-by-event fluctuations in particle multiplicity and radial flow \cite{Mazeliauskas:2015efa, Gardim:2019iah}. The PCA recently has been applied in unsupervised learning to test whether a machine can directly discover anisotropic flow coefficients from the high-volume simulation data without explicit instructions from human-beings \cite{Liu:2019jxg}. Because PCA can disentangle and extract the dominant components in data, it is a standard technique to perform dimensional reduction in statistical analysis as well as model emulation.

\textbf{Bayesian inference} or Bayesian analysis is based on the Bayes' theorem to derive posterior distributions of model parameters by constraining prior information with experimental data. It relies on a solid theoretical framework and precise experimental measurements. As already discussed in Sec.~\ref{Sec:viscosity}, it is a systematic way to constrain high-dimension model parameters using multiple experimental measurements. Over the past few years, Bayesian analysis has gradually become a standard tool to perform phenomenological extractions of the QGP properties from experimental measurements. The MADAI Collaboration initiated a community effort of applying Bayesian analysis to heavy-ion physics \cite{Novak:2013bqa}.  The Bayesian analysis has been applied to constrain the QCD Equation of State \cite{Pratt:2015zsa}, the QGP transport coefficients \cite{Sangaline:2015isa, Bernhard:2015hxa, Bernhard:2016tnd, Paquet:2017mny, Moreland:2018gsh, Bernhard:2019bmu, Paquet:2020rxl, Auvinen:2020mpc}, initial longitudinal fluctuations \cite{Ke:2016jrd}, QCD jet energy loss distribution \cite{He:2018gks, Soltz:2019aea}, and heavy-quark diffusion coefficients \cite{Xu:2017obm}. These works have paved a path towards precision physics in relativistic heavy-ion collisions. Bayesian analysis is a powerful tool to systematic extract information from experimental data using a well-established model. In the meantime, it has its limitation as the produced posterior distributions are influenced by the defined subjective priors. There is no unique and unbiased way to choose a prior in Bayesian analysis.

As an outlook, the next generation of Bayesian analysis in heavy-ion physics is evolving towards drawing global constraints with multiple subfields, such as combining bulk flow observables together with electromagnetic and QCD jet probes. Using unified theoretical frameworks and observables that probe multiple scales in heavy-ion collisions, more reliable and systematic information can be extracted from the Bayesian analysis.

\textbf{Deep Learning} (DL) is a branch of machine learning methods that are based on artificial neural networks to capture highly-correlated patterns/features in big data. It has achieved tremendous success in science and technology. The great advantage of the DL methods over conventional ones is its ability to extract hidden features from highly dynamical and complex nonlinear systems, like relativistic heavy-ion collisions.

The application of neural networks was pioneered in heavy-ion physics more than 20 years ago to determine the impact parameter of every heavy-ion collision based on final particle momentum distributions \cite{Bass:1996ez}. More recently, supervised learning with deep Convolutional Neural Networks (CNN) was used to identify the nature of QCD phase transition \cite{Pang:2016vdc, Du:2019civ}. The complex dynamics of heavy-ion collisions hide the experimental signals of a first-order phase transition and potentially enhanced fluctuations near a critical point.
Those works demonstrated that the deep CNN could provide a powerful and efficient ``decoder'' to extract information about QCD EoS from final particle momentum distributions. If the QGP fluid transits to hadron resonance gas through a first-order phase transition, the conserved net baryon density can clump together in space because of the spinodal decomposition. This phenomenon leaves characteristic imprints on the spatial net density distribution in every collision event, which can be detected by DL techniques \cite{Steinheimer:2019iso}. In the meantime, the impacts from spinodal decomposition on the measurable particle momentum space information are still challenging for DL methods to recognize.

DL techniques have also been applied to learn and mimic the nonlinear dynamics in relativistic heavy-ion collisions. A deep neural network was designed to learn and capture the main features of relativistic hydrodynamics \cite{Huang:2018fzn}. By treating initial energy density and flow velocity as inputs, the trained neural network can reproduce the realistic event-by-event hydrodynamical evolution on a quantitative level.  That work demonstrated that DL could speed up event-by-event simulations of heavy-ion collisions by orders of magnitude by replacing the real hydrodynamic simulations with the neural networks' predictions.

DL is gaining its popularity in a variety of aspects of heavy-ion physics to hunt for hidden experimental signals of important physics \cite{Chien:2018dfn, Lai:2018ixk, Komiske:2018cqr, Pang:2019aqb}. 

\section{Summary}

Relativistic heavy-ion collisions interconnect nuclear and high-energy physics. Experiments at the RHIC and the LHC push the field to evolve rapidly and bring us with many surprises along the way. The relativistic expanding Quark-Gluon Plasma is a unique fluid to study emergent many-body physics of strong QCD interactions. In the meantime, the QGP shares many universal collective features with other strongly-coupled systems in condensed matter and cosmology. Therefore understanding the properties of the QGP will not only advance our knowledge about the many-body aspects of strong interaction but also enrich the crosstalk with other fields in physics.

So far, theoretical studies on the out-of-equilibrium fluid dynamics in heavy-ion collisions are mostly limited to highly symmetric systems. These are QGP media experiencing either Bjorken expansion (0+1D), or Gubser expansion (1+1D), and being further simplified with respect to the conformal equation of state. In realistic heavy-ion collisions, however, neither of these symmetry conditions is rigorously satisfied. For instance, Bjorken boost invariance is apparently broken in proton-lead collisions. It is an inevitable step to examine the robustness of these attractors in realistic QGP in heavy-ion collisions, so that out-of-equilibrium hydrodynamics can provide a theoretical foundation for interpreting the collective behavior in small colliding systems.

Some generalized discussions have been carried out in various aspects. By numerically solving the second order viscous hydrodynamics, without Bjorken symmetry and with a non-conformal EoS, attractors are identified in multiple channels~\cite{Romatschke:2017acs}. In a similar manner, attractors from the kinetic theory solution without conformal symmetry is reproduced~\cite{Romatschke:2017acs,Florkowski:2017jnz}. In the context of kinetic theory, the $\L$-moments can be generalized in 3+1D expanding systems 
as well, by replacing the weight of Legendre polynomials by spherical harmonic functions. Attractors are found in the induced coupled moment equations~\cite{Kamata:2020mka}. These observations suggest the universal existence of attractors in out-of-equilibrium systems~\cite{Giacalone:2019ldn}.

Apart from the out-of-equilibrium extensions, the macroscopic fluid dynamics is an effective, robust, and efficient description of the bulk dynamics of heavy-ion collisions. It is a bridge that connects the fundamental QCD theory and experiments. Phenomenological studies combined with high precision flow measurements at RHIC and the LHC has been driving our field to a precision era. The adopting of the modern statistics and machine learning techniques, such as Bayesian analysis and Deep Learning, have become a popular and standard approach to extract the QGP transport properties systematically. The next challenge lies in how to quantify the model uncertainty in the theoretical framework. We are expecting more and more input from first principles calculations to reduce the theoretical uncertainties. 
Analytical solutions based on Bjorken and Gubser flow profiles have been widely adopted to validate hydrodynamic frameworks \cite{Marrochio:2013wla, Shen:2014vra, Noronha-Hostler:2014dqa, Bazow:2016yra, Pang:2018zzo}. In the recent years, there are a few formulations of relativistic causal hydrodynamics known as Israel-Stewart \cite{Israel:1979wp}, BRSSS \cite{Baier:2007ix}, DNMR \cite{Denicol:2012cn}, relativistic third-order dissipative hydrodynamics \cite{Jaiswal:2013vta}, and anisotropic hydrodynamics \cite{Bazow:2013ifa, Alqahtani:2017tnq, McNelis:2018jho}. These theories differ from each other by the number of velocity gradient terms included in the equations of motion for the dissipative tensors. Systematic comparisons among different hydrodynamics theories have been carried by comparing to the exact solution of the Boltzmann equation \cite{Florkowski:2013lza} under high degrees of symmetry \cite{Bazow:2013ifa, Florkowski:2014sfa, Denicol:2014mca, Florkowski:2014bba}. Future extensions of such comparisons to full (3+1)D will help our field to standardized the fluid dynamic model for relativistic heavy-ion collisions. A similar community-wide effort has been carried out for different transport frameworks in the field of low-energy nuclear physics \cite{Zhang:2017esm, Ono:2019ndq}.
Flow observables from the RHIC BES program and future FAIR/NICA experiments bring us to a new saga of full 3D dynamics beyond boost-invariant approximation. With the development of dynamical initialization schemes which interweave the 3D collision dynamics with fluid simulations, we are starting to quantify initial baryon stopping and study the collectivity of the QGP in a baryon-rich environment. This framework provides us with a reliable baseline to hunt for critical point signals in the upcoming RHIC BES II measurements. Flow measurements in small systems offers a window to study the early-stage dynamics of QGP. Understanding the collective origin in small systems has been leading the state-of-the-art theory development of rapid hydrodynamization as well as in- and out-of- equilibrium relativistic hydrodynamics.

\section*{Acknowledgments}
We thank J.~P. Blaizot, C. Gale, U. Heinz, , J.~F. Paquet, and B. Schenke for fruitful discussions. We thank the JETSCAPE Collaboration for providing preliminary results in Fig.~\ref{fig:viscosity_summary}. C.~S is supported in part by the U.S. Department of Energy (DOE) under grant number DE-SC0013460 and in part by the National Science Foundation (NSF) under grant number PHY-2012922. L.Y. is supported by National Natural Science Foundation of China under contract No. 11975079. This work is supported in part by the U.S. Department of Energy, Office of Science, Office of Nuclear Physics, within the framework of the Beam Energy Scan Theory (BEST) Topical Collaboration.

\bibliography{hydro_review_ref}

\end{document}